\documentclass[twocolumn]{aastex6}
\usepackage{array}
\usepackage{threeparttable} 
\newcolumntype{M}[1]{>{\centering\arraybackslash}m{#1}}

\usepackage{lineno}
\usepackage{float}
\fullcollaborationName{PALFA Collaboration}

\def\lapp{\ifmmode\stackrel{<}{_{\sim}}\else$\stackrel{<}{_{\sim}}$\fi}
\def\gapp{\ifmmode\stackrel{>}{_{\sim}}\else$\stackrel{>}{_{\sim}}$\fi}

\begin{document}

\title{PALFA Single-Pulse Pipeline: New Pulsars, Rotating Radio Transients, and a Candidate Fast Radio Burst}
\shorttitle{PALFA Single Pulse Discoveries}
\author{
C.~Patel\altaffilmark{1},
D.~Agarwal\altaffilmark{2},
M.~Bhardwaj\altaffilmark{1},
M.~M. Boyce\altaffilmark{1},
A.~Brazier\altaffilmark{3,4},
S.~Chatterjee\altaffilmark{5},
P.~Chawla\altaffilmark{1},
V. M.~Kaspi\altaffilmark{1},
D. R.~Lorimer\altaffilmark{2},
M. A.~McLaughlin\altaffilmark{2},
E.~Parent\altaffilmark{1},
Z.~Pleunis\altaffilmark{1},
S. M.~Ransom\altaffilmark{6},
P.~Scholz\altaffilmark{7},
R. S.~Wharton\altaffilmark{8},
W. W.~Zhu\altaffilmark{9,8,10},
M.~Alam\altaffilmark{11},
K.~Caballero Valdez\altaffilmark{12},
F.~Camilo\altaffilmark{13},
J. M. Cordes\altaffilmark{5},
F.~Crawford\altaffilmark{11},
J. S.~Deneva\altaffilmark{14},
R. D.~Ferdman\altaffilmark{15},
P. C. C.~Freire\altaffilmark{8},
J. W. T.~Hessels\altaffilmark{16,17},
B.~Nguyen\altaffilmark{11},
I.~Stairs\altaffilmark{18},
K.~Stovall\altaffilmark{19},
J.~van Leeuwen\altaffilmark{16,17}
}

\altaffiltext{1}{Department~of Physics and McGill Space Institute, McGill University, Montreal, QC H3A 2T8, Canada}
\altaffiltext{2}{Department of Physics and Astronomy, West Virginia University, Morgantown, WV 26506 \& Center for Gravitational Waves and Cosmology, West Virginia University, Chestnut Ridge Research Building, Morgantown, WV 26506}
\altaffiltext{3}{Department of Astronomy, Cornell Center for Astrophysics and Space Science, Space Science Building, Ithaca, NY 14853, USA and Cornell Center for Advanced Computing, Frank H.T. Rhodes Hall, Hoy Road, Ithaca, NY 14853, USA}
\altaffiltext{4}{Department of Astronomy, Cornell University, Ithaca, NY 14853, USA}
\altaffiltext{5}{Cornell Center for Astrophysics and Planetary Science and Department of Astronomy, Cornell University, Ithaca, NY 14853, USA}
\altaffiltext{6}{National Radio Astronomy Observatory, Charlottesville, VA 22903, USA}
\altaffiltext{7}{National Research Council of Canada, Herzberg Astronomy and Astrophysics, Dominion Radio Astrophysical Observatory, P.O. Box 248, Penticton, BC V2A 6J9, Canada}
\altaffiltext{8}{Max-Planck-Institut f̈ur Radioastronomie, Auf dem \"Hugel 69, D-53121 Bonn, Germany}
\altaffiltext{9}{National Astronomical Observatories, Chinese Academy of Science, 20A Datun Road, Chaoyang District, Beijing 100012, China}
\altaffiltext{10}{CAS Key Laboratory of FAST, NAOC, Chinese Academy of Sciences}
\altaffiltext{11}{Dept. of Physics and Astronomy, Franklin and Marshall College, Lancaster, PA 17604-3003, USA}
\altaffiltext{12}{University of Texas Rio Grande Valley, 1 W University Blvd, Brownsville, Tx 78520}
\altaffiltext{13}{SKA South Africa, Pinelands, 7405, South Africa}
\altaffiltext{14}{George Mason University, resident at the Naval Research Laboratory, Washington, DC 20375, USA}
\altaffiltext{15}{Faculty of Science, Univ. of East Anglia, Norwich Research Park, Norwich NR4 7TJ, United Kingdom} 
\altaffiltext{16}{ASTRON, The Netherlands Institute for Radio Astronomy, Postbus 2, 7990 AA, Dwingeloo, The Netherlands}
\altaffiltext{17}{Anton Pannekoek Institute for Astronomy, University of Amsterdam, Science Park 904, 1098 XH Amsterdam, The Netherlands}
\altaffiltext{18}{Dept. of Physics and Astronomy, University of British Columbia, Vancouver, BC V6T 1Z1, Canada}
\altaffiltext{19}{National Radio Astronomy Observatory, PO Box 0, Socorro, NM 87801, USA}

\altaffiltext{26}{Dept.~of Physics and Astronomy, West Virginia Univ., Morgantown, WV 26506, USA}
\altaffiltext{27}{Leibniz Universit ̈at Hannover, D-30167 Hannover, Germany}
\altaffiltext{28}{Max-Planck-Institut f̈ur Gravitationsphysik, D-30167 Hannover, Germany}
\altaffiltext{29}{Columbia Astrophysics Laboratory, Columbia Univ., New York, NY 10027, USA}
\altaffiltext{30}{Center for Advanced Radio Astronomy, Univ. of Texas Rio Grande Valley, Brownsville, TX 78520, USA }
\altaffiltext{31}{National Research Council, resident at the Naval Research Laboratory, Washington, DC 20375, USA} 
\altaffiltext{19}{Center for Gravitational Wave Astronomy, Univ. Texas - Brownsville, TX 78520, USA}
\altaffiltext{20}{Physics Dept., Univ. of Wisconsin - Milwaukee, 3135 N. Maryland Ave., Milwaukee, WI 53211, USA}
\altaffiltext{21}{Jodrell Bank Centre for Astrophys., School of Phys. and Astro., Univ. of Manchester, Manch., M13 9PL, UK}
\altaffiltext{24}{Dept. of Physics and Astronomy, Univ. of New Mexico, NM 87131, USA}

\begin{abstract}
We present a newly implemented single-pulse pipeline for the PALFA survey to efficiently identify single radio pulses from pulsars, Rotating Radio Transients (RRATs) and Fast Radio Bursts (FRBs). 
We have conducted a 
sensitivity analysis of this new pipeline in which multiple single pulses with a wide range of parameters were injected into PALFA data sets and run through the pipeline. 
Based on the recovered pulses, 
we find that for pulse widths $\rm < 5\ ms$ the sensitivity of the PALFA pipeline is at most a factor of $\rm \sim 2$ less sensitive to single pulses than our theoretical predictions. For pulse widths $\rm > 10\ ms$, as the $\rm DM$ decreases, the degradation in sensitivity gets worse and can increase up to a factor of $\rm \sim 4.5$.
Using this pipeline, we have thus far discovered 7 pulsars and 2 RRATs and identified 3 candidate RRATs and 1 candidate FRB. 
The confirmed pulsars and RRATs have DMs ranging from  133 to 386 pc~cm$^{-3}$ and
%
flux densities ranging from 20 to 160 mJy. The pulsar periods range from 0.4 to 2.1 s.
We report on candidate FRB 141113, which we argue is likely astrophysical and extragalactic, having $\rm DM \simeq 400\ pc~cm^{-3}$, which represents an excess over the Galactic maximum along this line of sight of $\rm \sim$ 100 - 200~pc~cm$^{-3}$.  We consider implications for the FRB population
and show via simulations that if FRB 141113 is real and extragalactic, the slope $\alpha$ of the distribution of integral source counts as a function of flux density ($N (>S) \propto S^{-\alpha}$) is $1.4 \pm 0.5$ (95\% confidence range). However this conclusion is dependent on several assumptions that require verification.

\end{abstract}

\section{Introduction} \label{sec:intro}
Pulsars are rapidly rotating, highly magnetized neutron stars (NSs). The majority of currently known pulsars are best detected through their time-averaged emission. Pulsar surveys like the PALFA survey \citep[Pulsar Arecibo L-band Feed Array;][]{cfl+06} generally use Fast Fourier Transform (FFT) searches in the frequency domain to search for pulsars. However, radio pulsar surveys often suffer from the presence of red noise generated by receiver gain instabilities and terrestrial interference.  This can reduce sensitivity, particularly to long-period pulsars. For example, \citet{lbh+15} reported that due to the presence of red noise, 
the sensitivity of the PALFA survey is significantly degraded for periods $\rm P > 0.5\ s$, with a greater  degradation in sensitivity for longer spin periods. In order to mitigate such problems, more effective time domain searches like the fast-folding algorithm ~\citep[FFA, see][and references therein]{lk05,kml+09,pkr+18} and single-pulse search techniques ~\citep[as described by][]{cm03} can be used. 

Rotating Radio Transients (RRATs) are a relatively recently discovered class of NSs that were detected only through their individual pulses~\citep{mll+06}. 
Due to the sporadic nature of their emission, surveys cannot rely on standard FFT searches to effectively look for RRAT signals. Instead, single-pulse search techniques 
are required.

Fast Radio Bursts (FRBs) are also a recently discovered phenomenon characterized by short (few $\rm ms$) radio bursts with high dispersion measures ($\rm DMs$)~\citep{lbm+07}. Unlike RRATs, which have observed $\rm DMs$ smaller than the maximum Galactic $\rm DM$ along the line of sight as predicted by Galactic free electron density models \citep{cl03,ymw16}, FRBs have $\rm DM$s that are much larger than this, implying extragalactic or even cosmological distances. To date, $\rm 34$ FRBs have been discovered\footnote{\url{www.frbcat.org}}, with only one FRB seen to repeat~\citep{ssh+16}. Like RRATs, FRBs can only be detected via single pulse-search techniques due to their transient nature. 

It is important to understand a survey's sensitivity to FRBs and RRATs as a function of various parameters (such as pulse width, $\rm DM$, scattering measure) if one is to accurately characterize the underlying sky event rates of these sources for population studies.

The PALFA Survey is the most sensitive wide-area survey for radio pulsars and short radio transients ever conducted.  Operating at a radio frequency band centered at $\rm 1.4\ GHz$, PALFA searches the Galactic plane ($\rm |b| < 5^{\circ}$), using the Arecibo Observatory, the 305-m single dish radio telescope located in Arecibo, Puerto Rico~\citep[see][for more details]{cfl+06,dcm+09,lbh+15}.  Since the survey began in $\rm 2004$, it has discovered $\rm 178$ pulsars, including 15 RRATs and one FRB.
~\cite{lbh+15} comprehensively characterized the sensitivity of PALFA to radio pulsars, and showed that it is sensitive to millisecond pulsars as predicted by theoretical models based on the radiometer equation which assumes white noise. However, PALFA suffers significant degradation to long-period pulsars due to the presence of red noise in the data. In order to improve the search for long-period pulsars, the PALFA collaboration has introduced a fast-folding algorithm~\citep{pkr+18}. 

\cite{dcm+09} described an early single-pulse search algorithm for PALFA, reporting on the discovery of seven objects.  
Here, we describe a new single-pulse search pipeline that we have also introduced to help identify long-period pulsars, RRATs and FRBs in our data. This new pipeline is described in \S\ref{sec:pipeline}. In \S\ref{sec:sensitivity}, we describe the survey's sensitivity to single pulses using an injection analysis. In \S\ref{sec:discoveries} we report new and candidate astrophysical sources discovered by this pipeline. We discuss a new candidate FRB, FRB 141113 in \S\ref{sec:frb}, and its implications for the FRB population in \S \ref{sec:implications}.  
We present our conclusions in \S\ref{sec:conclusion}.  

\section{The Single-Pulse Pipeline} \label{sec:pipeline}
\subsection{Overview of the pipeline} \label{sec:overview}
The PALFA survey uses a pipeline based on the software package $\tt{PRESTO}$ \citep{ran01} to search the observations for pulsars and radio transients. The processing is done on the Guillimin supercomputer which is the property of Compute Canada/Calcul Quebec, operated by McGill University's High Performance Computing Centre\footnote{\url{http://www.hpc.mcgill.ca/}}.

The data management, pre-processing of the data, Radio Frequency Interference (RFI) mitigation, dedispersion
 and single-pulse search techniques used by the PALFA consortium have been explained in detail by~\citet{dcm+09} and \citet{lbh+15}. 
Indeed, single-pulse searching has been a part of the pipeline since 2011. However, as described in this paper, the PALFA consortium has now implemented a more robust single-pulse pipeline in 2015 July. This required adding  more systematic and automated removal of radio frequency interference (RFI), as well as more automated candidate identification and visualization post-processing tools. After single-pulse searching using 
the standard {\tt PRESTO single\_pulse\_search.py} routine, the pipeline now makes use of a clustering algorithm to group single-pulse events and rank them (see \S\ref{sec:rrattrap}) according to a well defined metric to classify pulsars, RRATs and FRBs (henceforth ``astrophysical") candidates. A final diagnostic plot is produced for each candidate selected by the grouping algorithm so that it can be viewed by the members of the PALFA consortium to decide whether the candidate is astrophysical (see \S\ref{sec:waterfall} for more details). To aid in verifying astrophysical candidates, we introduced a series of heuristic ratings (\S\ref{sec:ratings}) and a machine-learning algorithm (\S\ref{sec:AI}) that is applied to each candidate. The candidates are viewed on an online collaborative facility, CyberSKA\footnote{\url{www.cyberska.org}}\citep[][\S\ref{sec:cyberska}]{kab+11}.        

\subsection{Grouping and Ranking of Single Pulses} \label{sec:rrattrap}
After the single-pulse search has been conducted on each time series, the output is sifted by the grouping algorithm \texttt{RRATtrap} \citep{kkl+15},  which clusters nearby single-pulse events into separate groups based on relative proximity in time and DM.  The grouped pulses are then ranked based on the criterion that the signal-to-noise ratio (S/N) of astrophysical pulses peaks at the optimal DM and falls off on either side (see top right plot in Fig.~\ref{fig:spd}). The \texttt{RRATtrap} algorithm was further improved and adapted for the PALFA survey as follows:
\begin{itemize}
    \item The relative proximity in DM and time between single-pulse events that is required to cluster them into a single group is now dependent on the DM, since the DM step size used for the search varies with DM \citep[see][]{lbh+15} instead of being fixed.  
    \item The minimum group size required for a cluster to  be considered a signal is no longer a fixed number but based on the expected S/N-DM curve 
    (Equations 12 and 13 of~\citet{cm03} and Equation~\ref{eq:cm03} below) given the observed S/N and pulse width.
    If the actual group size is smaller than the estimated one, the event is deemed to be noise.  
    \item If an astrophysical pulse is very narrow, it should only be detectable in  few neighboring DMs, with the number depending on the DM spacing. This results in a group size that is well below the minimum described above. In order to avoid missing these candidates, we created a new classification criterion. If the maximum S/N in the group of events is greater than 10, even if there are very few pulses ($<20$), for  a small pulse width ($<5$~ms) and a high DM ($\rm DM > 500\ pc\ cm^{-3}$),  this group is classified as astrophysical and is  subject to further investigation by the pipeline. 
    \item Pulses generated by narrow-band RFI tend to span a large $\rm DM$ range, but bright astrophysical pulses could also form groups that span large $\rm DM$ ranges. Instead of having a fixed number for a maximum allowed DM span as described by \citet{kkl+15}, we now estimate the DM range an astrophysical pulse should be detected over for a given S/N at the optimal DM as described by \citet{cm03}. If the group spans a DM range greater than a factor of five times our estimate, it is classified as RFI.
\end{itemize}   

\subsection{Production of the Single-Pulse Candidates} \label{sec:waterfall}
\begin{figure*}[h]
    \centering
    \includegraphics[width=1.0\textwidth]{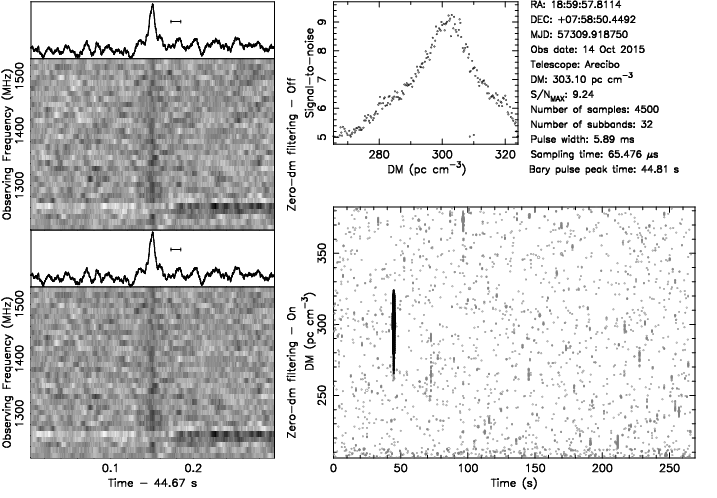}\\
    \vspace{1cm}
    \caption{
    `Spd' plot for RRAT J1859+07. \textit{Left} Dedispersed frequency vs time plots (dynamic spectra). The dedispersed time series are shown as line plots in the panels above the dedispersed frequency vs time greyscale plots, produced by summing the frequency channels below. The top plot is produced without zero-DM filtering, while the bottom plot uses a zero-DM filter \citep{ekl09}. \textit{Top Right:} S/N vs DM of the black pulse in the bottom right window showing that the S/N peaks near the optimal DM and decreases away from it. \textit{Bottom Right:} DM vs time for all single pulse events detected by $\tt{PRESTO's}$ $\tt{single\_pulse\_search.py}$. The higher the S/N of the event, the bigger the size of the point in the plot. The pulse in black is the candidate for which frequency vs time and S/N vs DM sub-plots plots are generated. Relevant header information about the candidate is displayed on the top right. See text for details.}
    \label{fig:spd}
\end{figure*}

In order to make the search process more efficient and systematic, all candidates classified as being astrophysical by the grouping algorithm (\S\ref{sec:rrattrap}) undergo automatic production of single-pulse diagnostic (`spd') plots to help with human verification. 
The spd plots contain all the features necessary to verify whether the candidate is astrophysical or is RFI. An example of such a plot is shown in Figure \ref{fig:spd} for RRAT J1859+07. 

On average, 20 such candidates are produced per beam. There is a binary output file produced for each candidate which can be used to reproduce the plot. These candidates are subject to a variety of heuristic ratings (see \S\ref{sec:ratings}) and exposed to a machine-learning algorithm that also rates them (\S\ref{sec:AI}). They are then uploaded to a database at the Center for Advanced Computing (CAC) located at Cornell University and can be viewed on our online candidate viewer (see Section~\ref{sec:cyberska}). 

\subsection{Ratings} \label{sec:ratings}
Currently, 10 heuristic ratings are applied to each single-pulse candidate produced by the pipeline, assessing different properties of the signal. The different ratings are described in Table~\ref{tab:ratings}. In the pipeline, they are applied to the candidate spd files. They assist the viewers in differentiating potential astrophysical candidates from RFI.              
\begin{table*}[b]

\small
\centering
\caption{Heuristic Single-Pulse Candidate Ratings}
\begin{tabular}{p{4cm} p{8cm}}                                                                                                                                              
\hline
\hline                                                                                                                                                                      
Ratings&Description\\ \hline
Peak Over RMS & The ratio of the peak amplitude of the profile to the RMS amplitude.\\
Phase Consistency & The fraction of sub-bands that are in-phase in the pulse window by $\rm \leq 2\%$. The offset is calculated by cross-correlating each interval with the summed profile.\\
Gaussian Amplitude & The amplitude of a single-Gaussian fit to the profile, normalized such that the profile standard deviation is 1.\\
Gaussian Goodness & The reduced $\chi^{2}$ of a single-Gaussian fit to the profile.\\
Gaussian FWHM & The full width at half-maximum of a single-Gaussian fit to the profile.\\
Fraction of Good Sub-bands & The fraction of frequency sub-bands above a set S/N threshold that contain the signal.\\
Sub-band S/N Standard Deviation & The standard deviation of the sub-band S/N ratios.\\
Known Pulsar Rating & The similarity of the position and DM to those of a known pulsar. The value is between 0 and 1, with values closer to 1 indicating similarity to a known pulsar.\\
Maximum $\rm DM$ Ratio & The ratio of the candidate $\rm DM$ to the maximum Galactic $\rm DM$ in the candidate direction according to the NE2001 electron density model~\citep{cl03}. 
\\
\hline                                                                                                                                                     
\end{tabular}
\label{tab:ratings}
\end{table*} 
\subsection{Machine Learning Candidate Selection} \label{sec:AI}
The single-pulse candidates produced by the pipeline are also exposed to a machine-learning algorithm that
attempts to select astrophysical candidates. The single-pulse pipeline uses the same machine-learning algorithm as employed by the periodicity pipeline and explained by \citet{zbm+14}. It uses an image pattern recognition system that mimics humans to distinguish pulsar signals from noise/RFI candidates. The algorithm is trained regularly based on the manual classification of candidates by members of the collaboration. Since the algorithm was designed to view the periodicity 
candidate plots, slight changes were made for it to work on the single-pulse `spd’ candidate plots (Figure~\ref{fig:spd}):
\begin{itemize}
    \item The zero-DM filtered, dedispersed time series of the `spd' plot replaces the pulse profile of a periodicity candidate;
    \item The zero-DM filtered, dedispersed dynamic spectrum of the `spd' plot acts like time vs phase and frequency vs phase sub-plot of a periodicity candidate; and
    \item The dynamic spectrum of the `spd' plot is dedispersed for a range of DMs around the best DM, and a time series is produced for each DM trial. The time series for each DM is analyzed and a plot of reduced $\chi^{2}$ vs DM (similar to that of a periodicity candidate) is produced for the machine learning algorithm to analyze.
\end{itemize}

\subsection{Candidate Viewer: CyberSKA}
\label{sec:cyberska}
All the candidates produced by our pipeline are uploaded to the results database at CAC.  The results can be viewed online via the CyberSKA portal \citep{kab+11}. The PALFA collaboration has developed several applications on this portal for viewing periodicity search candidates \citep{lbh+15}.  We developed a new application for viewing single-pulse candidates that is very similar to the existing application.
Specifically, we can filter using queries on different candidate properties, ratings (\S\ref{sec:ratings}) and file metadata information. As with our periodicity candidate viewer, single-pulse candidates can be classified as astrophysical, RFI, noise or known sources. The best candidates are uploaded to a Top Candidates database and are eventually followed-up for confirmation.  

\section{Survey Sensitivity to Single Pulses}
\label{sec:sensitivity}
The peak flux density of single pulses are generally estimated using the following equation from ~\citet{cm03}:
\begin{equation}\label{eq:cm03}
    \rm S_{i} = \frac{\beta(S/N)_{b}(T_{sys}+T_{sky})}{GW_{i}}\sqrt{\frac{W_{b}}{n_{p}\Delta f}},
    \label{eq:cl03}
\end{equation}
where $\rm S_{i}$ is the intrinsic flux density, $\rm \beta$ is a factor accounting for the sensitivity loss due to digitization, $\rm (S/N)_{b}$ is the signal-to-noise ratio of the broadened pulse, $\rm T_{sys}$ and $\rm T_{sky}$ are the system temperature at the observing frequency and the sky temperature, respectively, $\rm G$ is the telescope gain, $\rm W_{i}$ and $\rm W_{b}$ are the intrinsic and broadened pulse widths, respectively, $\rm n_{p}$ is the number of summed polarizations, and $\rm \Delta f$ is the observing bandwidth. 
Equation~\ref{eq:cm03} is a theoretical representation of the sensitivity to single pulses in the presence of Gaussian noise. The sensitivity to single pulses in real survey data (which contains RFI and other non-Gaussian features) can be significantly different from the theoretical estimates. Here, we describe an injection analysis to better characterize our survey's sensitivity. 

\subsection{Injection of Single Pulses}
\label{sec:injection}
We used the same data set ($\rm 12$ distinct and calibrated observations) that was used by~\citet{lbh+15} and injected synthetic signals into those observations as previously described. A pulse was injected every $\rm \sim 10\ s$ yielding $\rm 26$ pulses per observation (of duration 268 s). In a single observation, all the injected pulses had the same parameters (i.e. pulse width, DM and amplitude). Since the data quality can vary during an observation due to RFI, our method helps us characterize our sensitivity over the entire observation and provides a large statistical sample of pulses from which to draw conclusions. 
Even though the injected pulses within an observation had the same parameters, we repeated the process with a new set of parameters which allowed us span a wide range of pulse characteristics (see Table~\ref{tab:injectionparams}) for our analysis. In order to vary the pulse width in the injection algorithm used by~\citet{lbh+15}, all pulses were injected using the same duty cycle of $\rm \sim$1.5$\%$, but with different pulse periods. For the first set of injection trials, the injected pulses were not subject to scatter-broadening.  In the second set of injections, we fixed the DM and pulse width and varied scattering times. 

\begin{table}
\small
\centering
\caption{Injected Pulse Parameters}
\begin{tabular}{p{3.5cm}p{4cm}}
\hline
\hline
Parameter & Values\\
\hline
Pulse Width ($\rm ms$) & $\rm \sim 1,\ 2,\ 5,\ 10,\ 20$\\ 
$\rm DM$ ($\rm pc\ cm^{-3}$) & $\rm 42.5,\ 153.7,\ 326.2,\ 615.5$\\
&$\rm 1005.7,\ 5606.7$\\
Scattering Time ($\rm ms$) & ... \\
\hline
\hline
Pulse Width ($\rm ms$) & $\rm \sim5$\\
$\rm DM$ ($\rm pc\ cm^{-3}$) & $\rm 1005.7$\\
Scattering Time ($\rm ms$) & $\rm 10,\ 20,\ 50,\ 100$\\
\hline
\end{tabular}
\label{tab:injectionparams}
\end{table}

\subsection{Results of Sensitivity Analysis}
\label{results_sensitivity}

\begin{figure*}[h]
    \centering
    \includegraphics[width=1.0\textwidth]{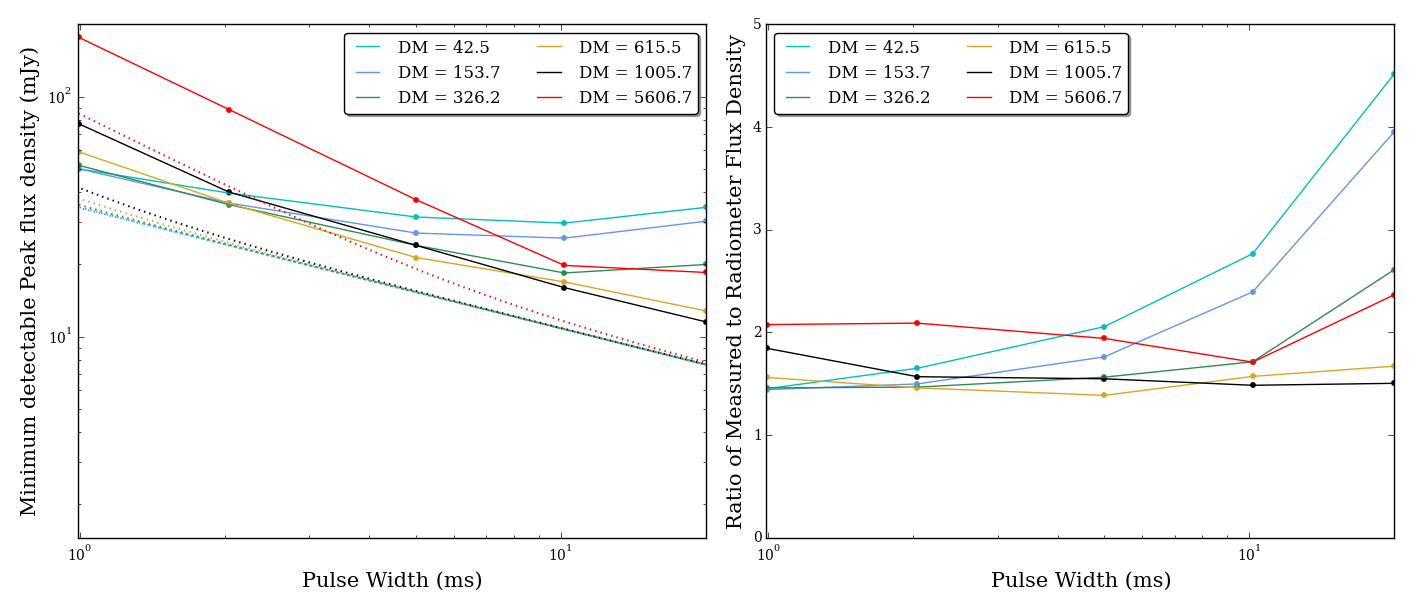}
    \caption{PALFA's sensitivity to single pulses in the absense of scattering. Plotted is the minimum detectable peak flux density as a function of pulse width in the absence of pulse broadening due to scattering (left) and the ratio of the measured flux density limit to the predicted flux density limit (right). The dotted lines show the theoretical predictions as given by Equation~\ref{eq:cm03} and the points show the actual detections where $\rm >90\%$ of the injected pulses with $\rm S/N>7$ were recovered by the pipeline. The points have been connected by straight lines to guide the eye. The different colors correspond to different $\rm DMs$ ranging from $\sim$42--5600~pc~cm$^{-3}$. The sensitivity curves in the right-hand plot show that for pulse widths $\rm < 5\ ms$ our survey is at most a factor of $\rm \sim 2$ less sensitive to single pulses than the theoretical predictions. For pulse widths $\rm > 10\ ms$, as the $\rm DM$ decreases, the relative degradation in sensitivity can increase up to a factor of $\rm \sim 4.5$.}
    \label{fig:sensitivity_noscatt}
\end{figure*}
\begin{figure*}[h]
    \centering
    \includegraphics[width=1.0\textwidth]{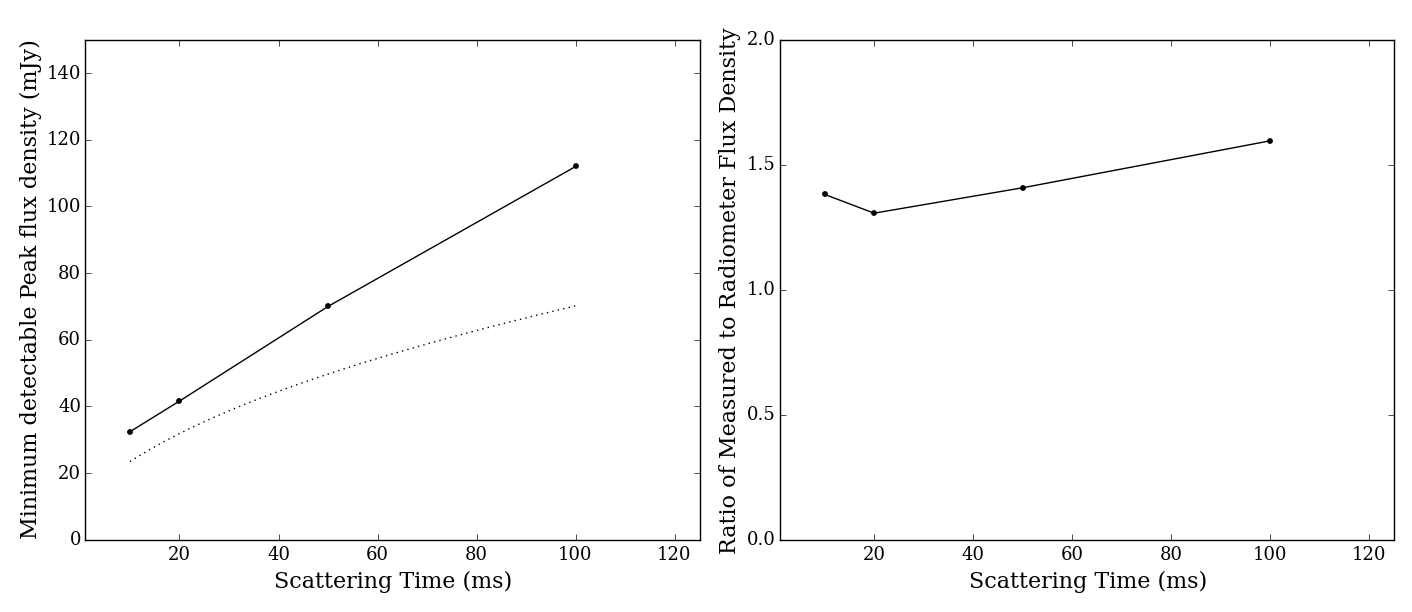}
    \caption{PALFA's sensitivity to single pulses in the presence of scattering for DM = 1005.7~pc~cm$^{-3}$.  We plot  the minimum detectable flux density as a function of scattering times when pulse broadening is dominated by scattering (left) and
    its ratio to the theoretical prediction (right) based on Equation~\ref{eq:cm03}. The injections were done for a single DM and a single intrinsic pulse width. The dotted line is the theoretical prediction given by Equation~\ref{eq:cm03} and the solid line is the actual `detection' limit, for which $>90\%$ of the injected pulses with $\rm S/N>7$ were recovered by the pipeline. The curve shows that for this set of parameters, the survey's sensitivity is degraded by less than  a factor of 2 from the theoretical prediction.}
    \label{fig:sensitivity_withscatt}
\end{figure*}

The data with injected pulses were processed by the single-pulse pipeline described in \S\ref{sec:pipeline}. Every pulse in a single observation was injected with an amplitude corresponding to an initial best guess for the limiting flux density. 
The output of the pipeline was classified as either a detection or a non-detection.  Since all injected pulses in a single observation were given the same amplitude, the pipeline output was classified as a detection if at least 24 of the 26 injected pulses in a single observation were successfully detected with $\rm S/N > 7$, giving us $\rm > 90\%$ confidence of detecting a single pulse above $\rm S/N$ of $\rm 7$. In this case, we reduced the amplitude by 20\% for the next injection trial. In case of a non-detection, the flux of the single pulses was increased by $\rm 20\%$ for the next injection trial. The injected flux was varied in this way until the difference between the fluxes of outputs classified as a `detection' and a `non-detection' was less than $10 \%$. The injected flux at this point was assigned to be the sensitivity limit for the corresponding set of injection parameters and the observation used the median of the sensitivity limits from all 12 observations was declared to be the survey's minimum detectable peak flux density for the corresponding set of injection parameters (i.e, $\rm DM$, pulse width, and scattering time).

We show the results of the sensitivity analysis in the absence of scattering in Figure~\ref{fig:sensitivity_noscatt}. As expected from the theoretical predictions, the minimum detectable flux density increases with DM for all pulse widths. For all DMs, it decreases as the pulse width increases. The comparison between the measured and the theoretical sensitivities of the survey is shown in the right plot of Figure~\ref{fig:sensitivity_noscatt}. For low pulse widths ($\rm < 5\ ms$), the survey suffers a degradation in sensitivity by a factor of $\rm \sim 1.5$ at low DMs to $\rm \sim 2$ at high DMs, compared to the theoretical estimates. For low pulse widths and high DMs, this loss in sensitivity is primarily due to intra-channel smearing. We also find that for large pulse widths ($\rm > 5\ ms$), as the $\rm DM$ decreases, 
the degradation in sensitivity increases to a factor of $\sim 4.5$ from the theoretical predictions. We understand that this can be attributed to zero-DM filtering \citep{ekl09} that we perform to mitigate broadband terrestrial RFI. While this technique is excellent at mitigating terrestrial RFI, it also removes power from astrophysical signals at low DMs.

Next we introduced scattering to the injected pulses assuming $\rm DM = 1005.7$~pc~cm$^{-3}$ and a pulse width of 5~ms in the same data set. For PALFA pointings, scattering timescales at high DMs can range from a few microseconds (if pointing at high Galactic latitudes toward the outer Galaxy) to a few seconds (for low Galactic latitudes towards the inner Galaxy). Since PALFA does not search for $\rm pulse\ widths > 100\ ms$, the scattering time scales for this analysis were less than $\rm 100\ ms$. The injection parameters are shown in Table~\ref{tab:injectionparams}. Again, 26 pulses were injected for a single trial with  a detection declared if at least 24 pulses with $\rm S/N > 7$ were recovered by our single-pulse pipeline. The results of this analysis are shown in Figure~\ref{fig:sensitivity_withscatt}. The sensitivity of our survey to single pulses for these parameters is a factor of $\rm \sim 1.5$ lower than that predicted by Equation~\ref{eq:cm03}. This is roughly the same amount of degradation that we find for single pulses (pulse widths $\rm 5 \ ms$ and $\rm DM \sim 1000$~pc~cm$^{-3}$) that are not subject to scattering (see Fig.~\ref{fig:sensitivity_noscatt}, right).
This indicates that Equation~\ref{eq:cm03} adequately models the effects of scattering.

\section{New Discoveries and Candidates} \label{sec:discoveries}

\begin{figure*}[ht!]
      \centering
          \includegraphics[width=0.45\textwidth]{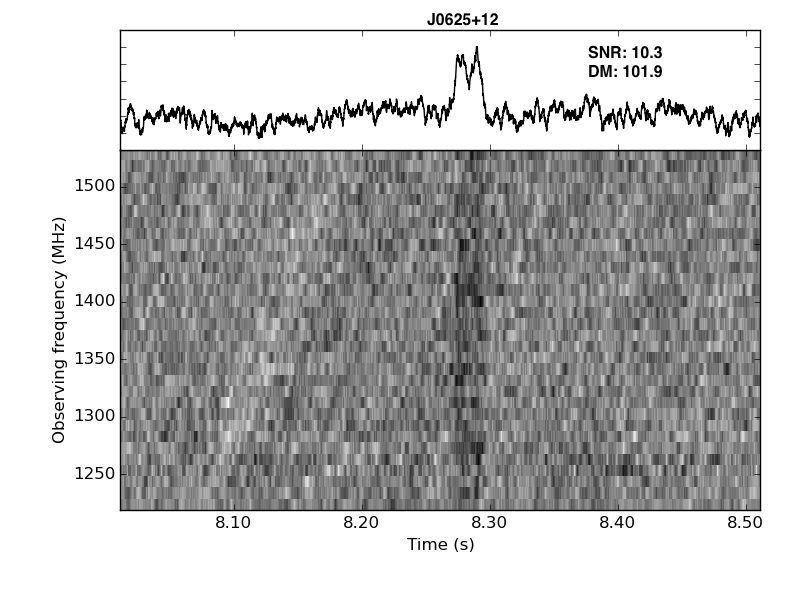}
           \includegraphics[width=0.45\textwidth]{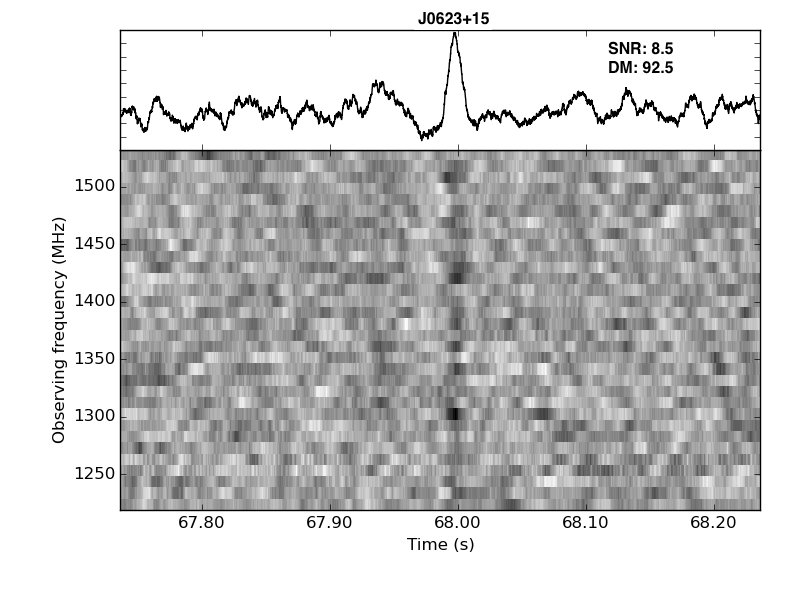}
            \includegraphics[width=0.45\textwidth]{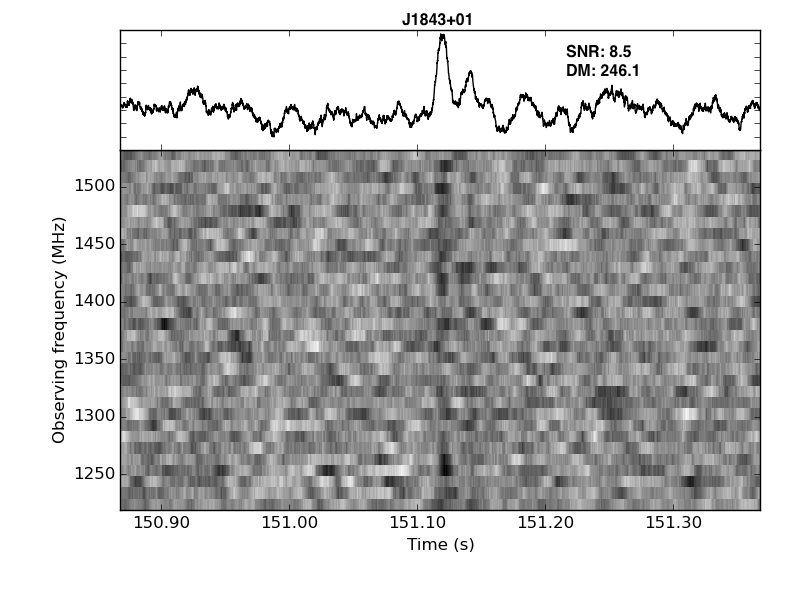}
            \includegraphics[width=0.45\textwidth]{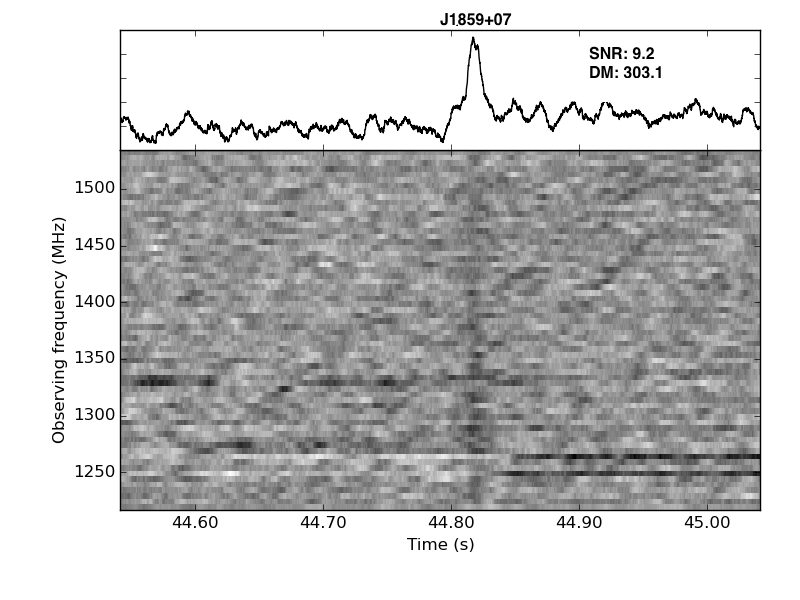}
  \caption{Dedispersed frequency vs time plots for pulsars and RRATs discovered by the single pulse pipeline. The instrument bandpass has been subtracted.}
    \label{fig:waterfall_plots}
\end{figure*}

The new single-pulse pipeline has been fully incorporated into our main data analysis pipeline since 2015 July, during which we have processed  $\sim$60,500 beams as of 2018 February 10. With 7 beams per PALFA survey pointing and with each being 268-s long in the inner Galaxy and 180-s long in the outer Galaxy, we have just under 24 days of total observing time. From the number of beams processed, this pipeline has reported a total of $\sim$900,000 single-pulse candidates (grouped single pulses). Out of these, $\sim$55,000 single-pulse candidates have been classified by members of the PALFA collaboration with our web viewer (\S\ref{sec:cyberska}) using a variety of filters and ratings (\S\ref{sec:ratings}). Of the classified candidates, $\sim$46,000 have been classified as being RFI or noise, $\sim$3,800 as potential astrophysical candidates and $\sim$4,900 as known astrophysical sources. The single-pulse pipeline has uniquely discovered 3 pulsars (2 RRATs and 1 pulsar). Additionally, it has independently discovered 6 pulsars which were also detected using our standard periodicity analysis. It has also identified 3 candidate RRATs and one candidate FRB (see \S\ref{sec:frb}). The details of the new discoveries are presented in Table~\ref{tab:discoveries} and their dedispersed frequency vs time plots are shown in Figure \ref{fig:waterfall_plots}. The $\rm w_{50}$ (full width at half maximum) and $\rm w_{90}$ (full width at a tenth of the maximum) pulse widths for each discovery candidate were estimated by fitting a Gaussian to their pulse profiles. In order to estimate the peak flux densities reported in Table~\ref{tab:discoveries}, we used the radiometer equation (Equation~\ref{eq:cm03}), for which we used $\rm T_{sys}+T_{sky} = 30\ K$, telescope gain $\rm G = 8.2\ K/Jy$ ~\citep{sch+14}, $\rm \beta = 0.9$, $\rm n_{p} = 2$ and $\rm \Delta f = 322\ MHz$. The $\rm S/N$ and $\rm w_{90}$ (used as $\rm W_{b}$, the broadened pulse width) were taken from Table~\ref{tab:discoveries}. For each source, we estimated the degradation factor by choosing the right-hand curve in Figure~\ref{fig:sensitivity_noscatt} corresponding to the nearest DM value, and the factor (ratio of measured to radiometer flux density limit) corresponding to its pulse width. We then applied the degradation factor to the peak flux density estimated by the radiometer equation. 
 
All the discoveries in the upper section of Table~\ref{tab:discoveries} have been confirmed via re-observations and are now being monitored by either the Lovell Telescope at Jodrell Bank Observatory or with the Arecibo Observatory as a part of our timing campaign. Their detailed timing properties will be reported upon in a future publication.

\begin{table*}[h]
\small
\centering
\caption{New Discoveries from the Single-Pulse Pipeline}
\begin{tabular}{M{2.5cm}M{2.5cm}M{2.0cm}M{0.01cm}M{2.5cm}M{2.cm}M{1.cm}M{2.cm}M{2.cm}}
\hline
\hline
Name& Detection method& Pulse width (ms)& & Period (ms)& $\rm DM\ (pc\ cm^{-3})$& S/N& Degradation Factor& Flux Density (mJy)\\
\end{tabular}
\begin{tabular}{M{2.5cm}M{2.5cm}M{1.cm}M{1.cm}M{2.5cm}M{2.cm}M{1.cm}M{2.cm}M{2.cm}}
&&$\rm w_{50}$&$\rm w_{90}$&&&&&\\
\hline
PSR J1859+07& SP& 4.5& 8.1& ...& $\rm 303.1\pm 2.2$& 9.2& 1.5& 20\\
PSR J1905+0414& SP& 3.3& 5.9& ...& $\rm 383 \pm 1$& 14.2& 1.5& 36\\
PSR J1952+30 & SP& 5.7& 10.5& $\rm 1665.60\pm 0.12$& $\rm 188.8\pm 0.6$& 10.4& 2.5& 33\\
PSR J1856+09& SP \& Periodicity& 4 & 7.3& $\rm 2170.71\pm 0.11$& $\rm 193.4\pm 0.6$& 15.6& 2& 48\\
PSR J1853+04& SP \& Periodicity& 2& 3.8& $\rm 1320.65\pm 0.04$& $\rm 549.3\pm1.3$& 10.0& 1.5& 33\\
PSR J1958+30& SP \& Periodicity& 4& 7.3& $\rm1098.53\pm0.02$& $\rm 199.3\pm 0.4$& 17.5& 2& 54\\
PSR J2000+29& SP \& Periodicity& 7.4& 13.5& $\rm 3073.70\pm 0.14$& $\rm 132.5\pm 1.4$& 13.1& 3& 159\\
PSR J1901+11& SP \& Periodicity& 2.2& 4& $\rm 409.14\pm 0.01$& $\rm 268.9\pm 0.8$& 9.2& 1.5& 29\\
PSR J1843+01 & SP \& Periodicity& 3.5& 6.4& $\rm 1267.02\pm0.04$& $\rm 247.8\pm2.4$& 8.5& 1.5& 21 \\

\hline

Candidate PSR J0625+12& SP& 7.1& 12.9& ...& $\rm 101.9\pm 6.1$& 10.3& 3& 36\\
Candidate PSR J0623+15& SP& 14.1& 25.7& ...& $\rm 92.5\pm 1.6$& 8.5& 4.5& 32\\
Candidate PSR J1908+13& SP& 5.1& 9.2& ...& $\rm 180.3\pm 1.1$& 19.2& 2& 52\\ 
\hline
Candidate FRB 141103& SP& 1.1& 2& ...& $\rm 400 \pm 3$& 8.4& 1.5& 39\\
\hline
\hline
\end{tabular}
 \begin{tablenotes}
   \small
   \item $^\diamond$ The flux densities were calculated assuming that the pulses were detected in the center of the beams. 
  \end{tablenotes}
\label{tab:discoveries}
\end{table*}

\begin{figure*}[hb!]
    \centering
    \includegraphics[width=0.45\textwidth]{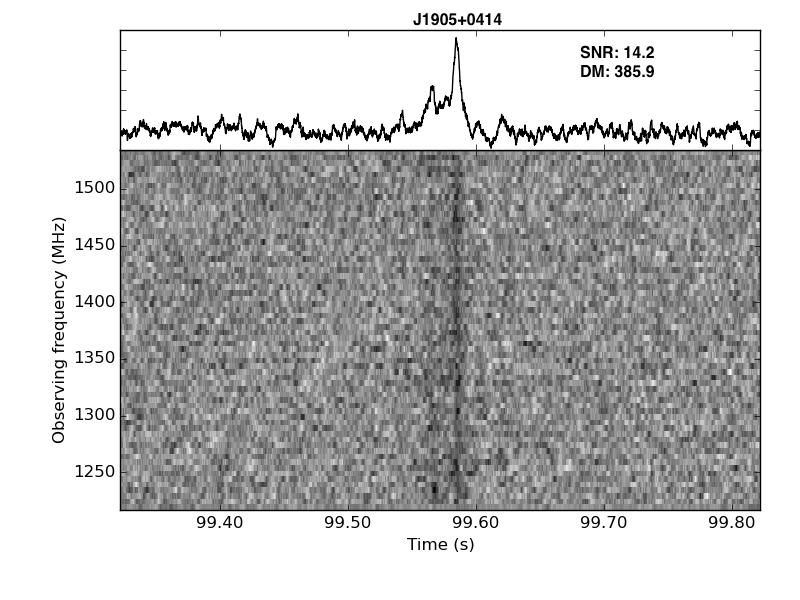}
    \includegraphics[width=0.45\textwidth]{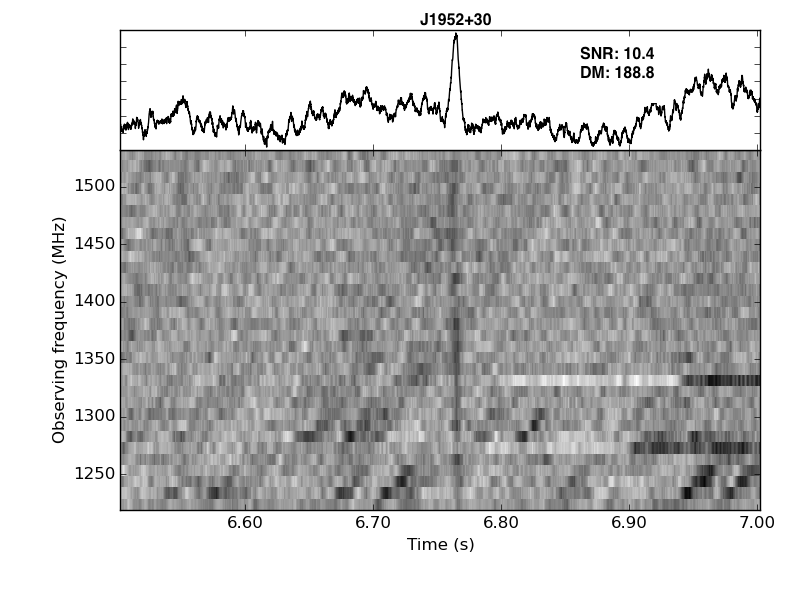}
      \includegraphics[width=0.45\textwidth]{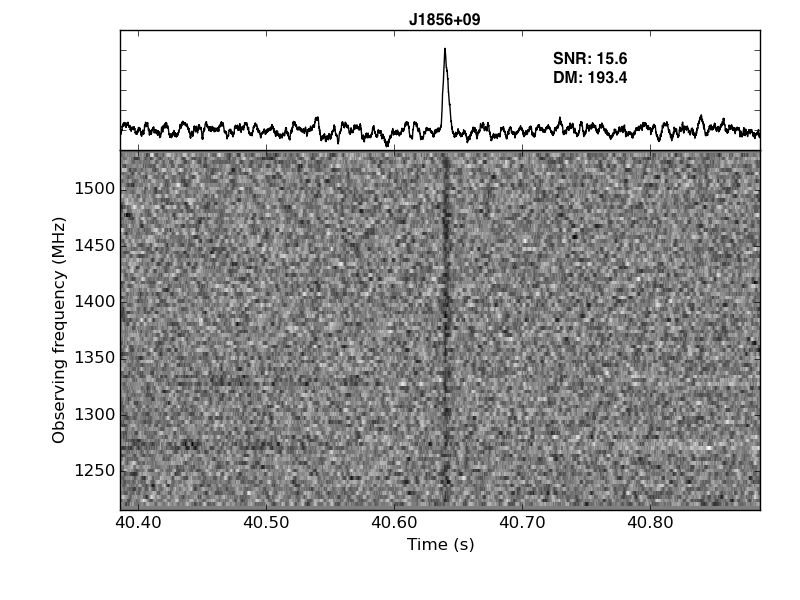}
       \includegraphics[width=0.45\textwidth]{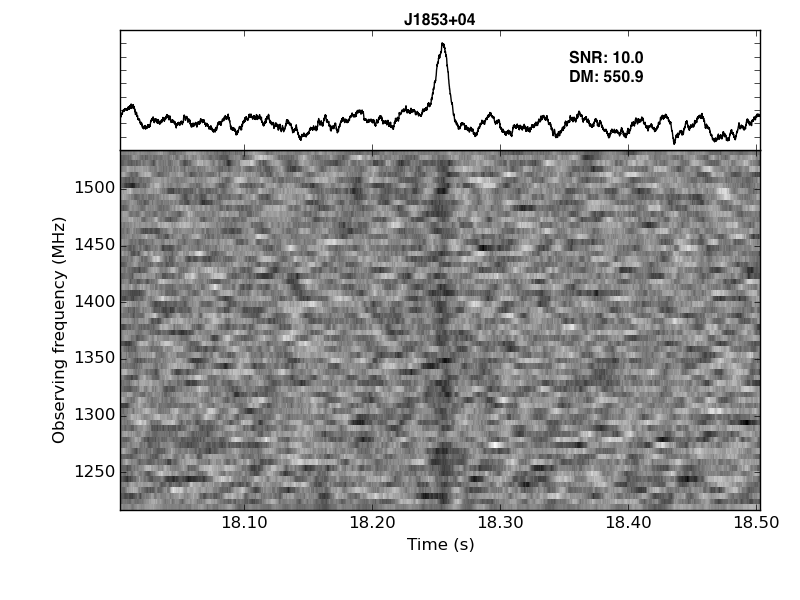}
         \includegraphics[width=0.45\textwidth]{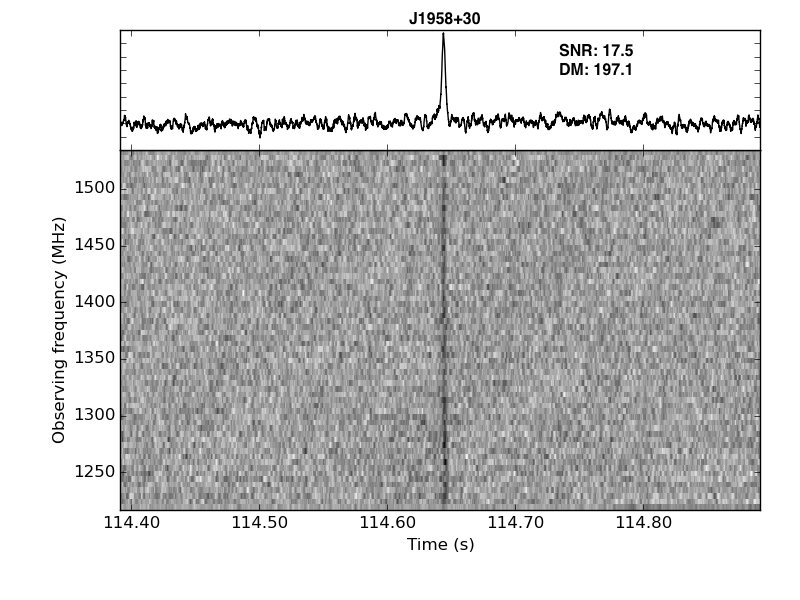}
         \includegraphics[width=0.45\textwidth]{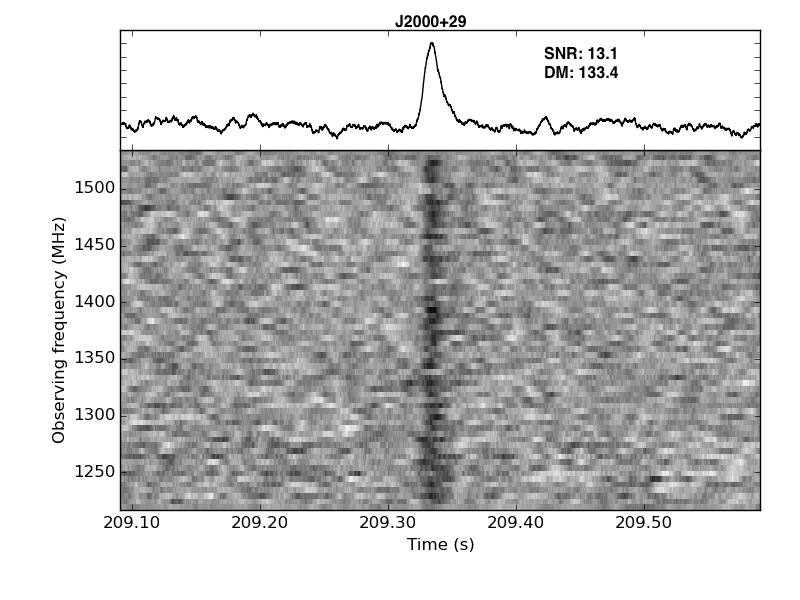}
         \includegraphics[width=0.45\textwidth]{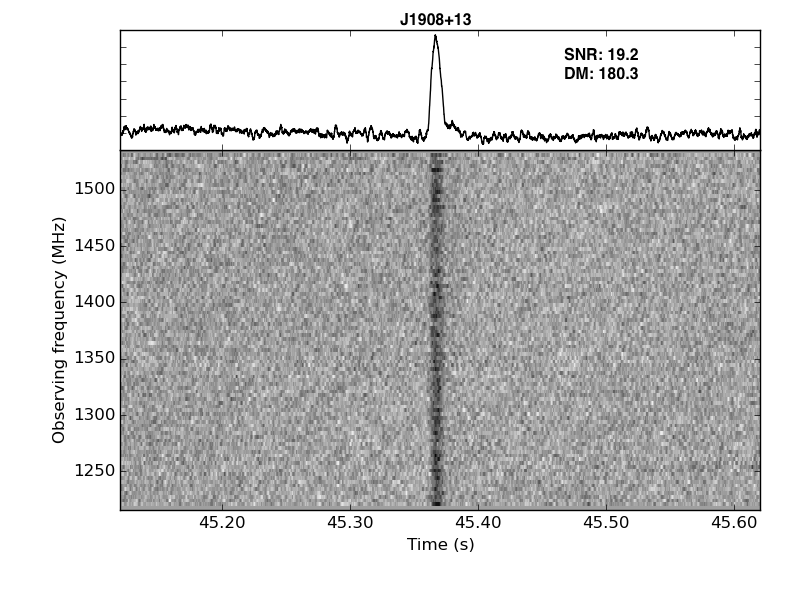}
         \includegraphics[width=0.45\textwidth]{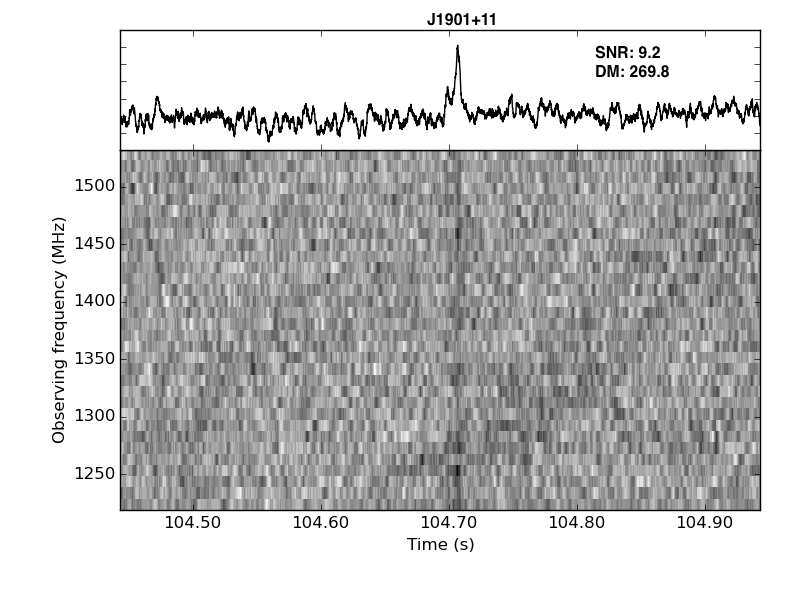}
 
\end{figure*}

\section{Population synthesis of RRATs in the PALFA survey}\label{potentialrrats}

To predict the number of RRATs that should have been detected by the survey to date we follow the approach for the RRAT population model developed by \citet{agarwal2018} who have recently adapted the pulsar population software \textsc{PsrPopPy2}\footnote{https://github.com/devanshkv/PsrPopPy2} \citep{b8s2014} to model the Galactic population of RRATs.  In their optimal model for the underlying RRAT population, when passed though model surveys, the resulting model-detected population closely resembles the observed RRAT population.

We developed a population model based on RRATs detected by four surveys: the Parkes multibeam survey \citep{PM1,k11,pmsurv}, the high-time-resolution intermediate survey \citep{HTRU-RRATs, HTRUmid}, and two higher latitude surveys done with the Parkes radio telescope and reported in \citet{BS10, Jacoby, Edwards}.
We follow a method similar to the method used by \cite{Lorimer2006} for constructing a ``snapshot'' (i.e.~no time evolution) of the underlying  RRAT population. We begin with uniform underlying distributions for the period,  luminosity, Galactocentric radius, and  burst rate. We use an exponential
distribution for the Galactic $Z$ distribution with a mean scale height of 0.33~kpc (as for  pulsars). A total of 11,000 RRATs are drawn with these distributions and run through surveys mentioned above. This number is set much higher than the actual number detected through the surveys above to minimize statistical fluctuations. The model detected population is then compared with the RRATs detected from these surveys by calculating the reduced $\chi^2$ of the distributions in $R$,
$L$, $Z$ and burst rate. As described by \cite{Lorimer2006}, correction factors are applied to the underlying population to refine the models, and the process is repeated until the reduced $\chi^2$ between the observed and detected model population is $\sim$ 1. Full details of this analysis are given in  \citet{agarwal2018}.

\begin{figure*}[t]
\centering
\includegraphics[width=0.85\textwidth]{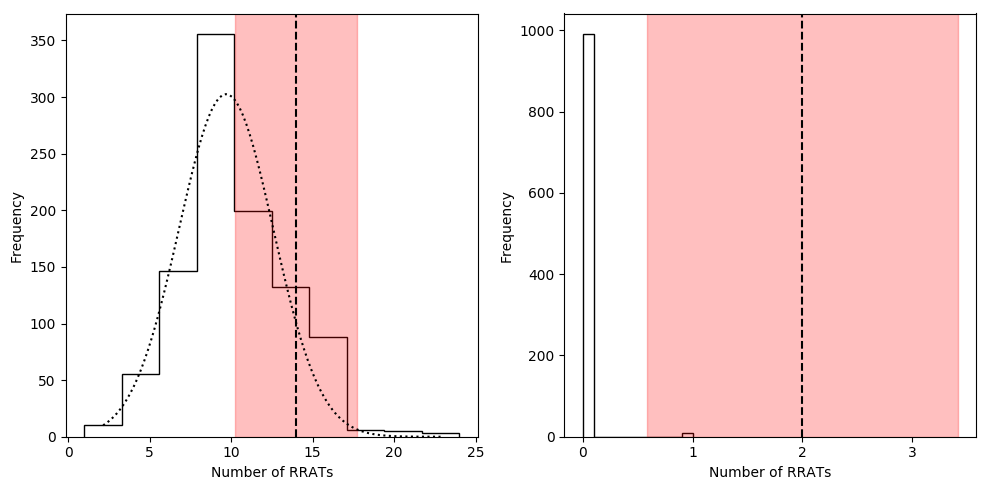}
\caption{Distribution of the expected number of RRATs detected with the inner (left) and outer (right) Galaxy PALFA survey to date for 1000 simulations. The dashed vertical line represents the number of RRATs detected so far in each survey. The red shaded region represents Poissonian  $\sqrt{N}$ uncertainties in the number of detected RRATs. The dotted line in the left plot shows the Gaussian fit for the distribution for the inner Galaxy survey (see text).}
\label{fig:palfa_pred}
\end{figure*}

Using the optimal model from this procedure, we generate a population such that it detects 55 RRATs in the four surveys. We then run the inner and outer galaxy PALFA surveys to find the number of RRATs detected. 
This process is repeated 1000 times to get a distribution of the number of RRATs detectable by our survey.
Figure ~\ref{fig:palfa_pred} shows the distribution of the number of detected RRATs by our inner Galaxy survey. The distribution is well fit by a Gaussian with mean $\mu = 9.6 \pm 0.3$ and standard deviation $\sigma=2.8 \pm 0.3$.

As shown in Figure~\ref{fig:palfa_pred}, this 
simulation is in good agreement with the number of RRATs we find for the inner Galaxy survey. The same procedure was repeated for the outer galaxy survey which and from this we predict zero detections. This appears to be in tension with the fact that PALFA has so far found
two RRATs in the outer Galaxy
survey. Although partially attributable to small-number statistics, the discrepancy could be an indication that the population model is biased towards RRATs in the inner Galaxy. The four surveys used in constructing the model targeted the inner Galaxy. In the future, we will use discoveries from the PALFA survey to construct an improved RRAT population model, taking into account discoveries in the outer Galaxy.


\section{Candidate FRB 141113}
\label{sec:frb}
\begin{figure*}[hbt]
    \centering
    \includegraphics[width=0.7\textwidth]{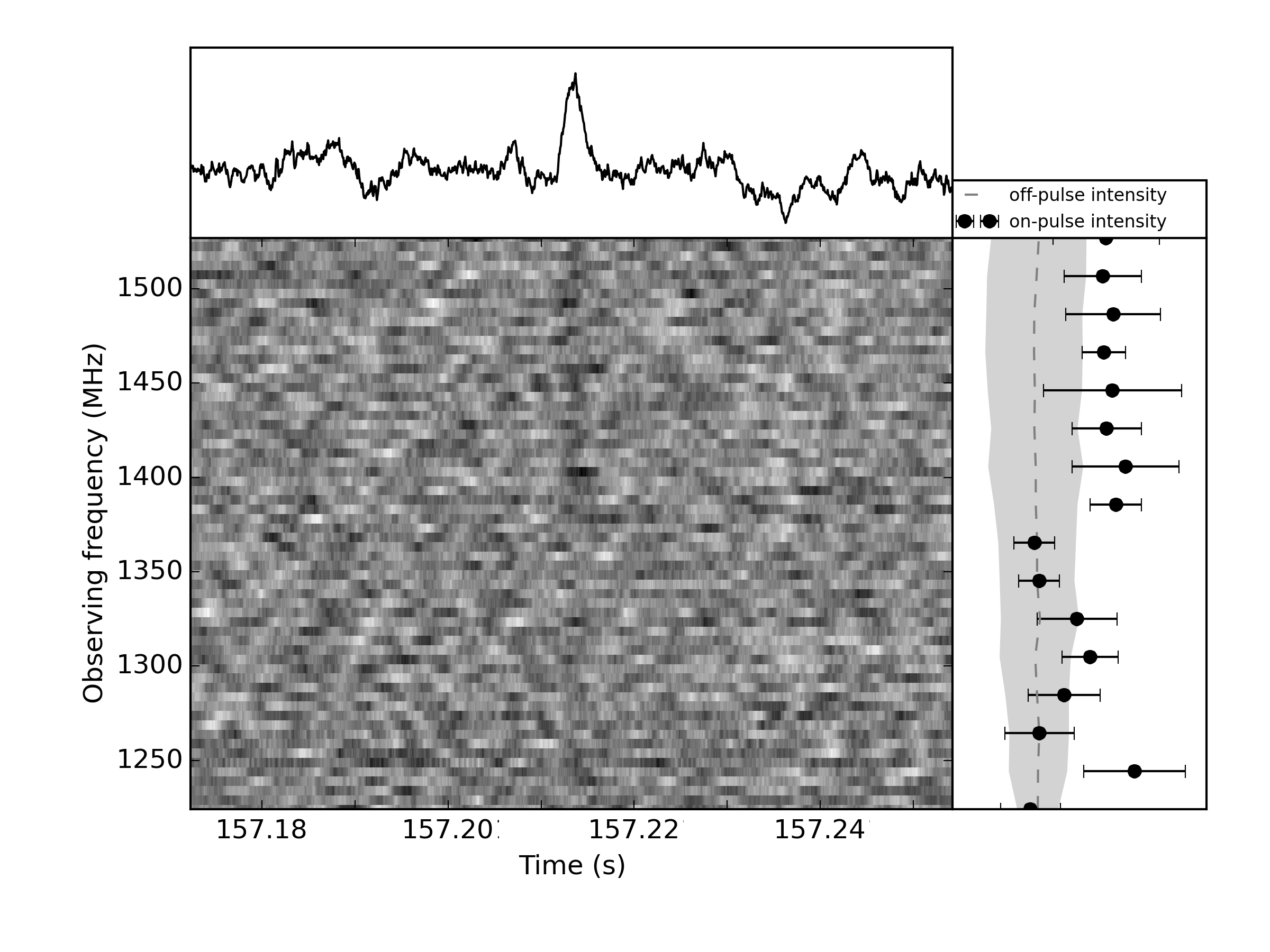}
    \caption{Dedispersed frequency vs time plot for the candidate FRB 141113. 
             The pulse was detected with $\rm S/N = 8.4$, pulse width 
             $\rm \sim 2\ ms$, and $\rm DM=400\ pc\ cm^{-3}$ (well above the Galactic contribution). For clarity, we used a frequency resolution of 64 sub-bands. The instrumental bandpass has been subtracted. On the right are mean on-pulse (scatter points) and off-pulse (dashed line) relative intensities binned over 16 sub-bands. The on-pulse area was taken to be the $\rm W_{90}$ (from Table~\ref{tab:discoveries}) region around the peak of the pulse. The remainder was considered as off-pulse region. The gray band shows the $\rm 1\sigma$ range around the mean of the off-pulse intensity. The error bars on the scatter points indicate $\rm 1\sigma$ spread around the mean of the on-pulse intensity.}
    \label{fig:0613+18}
\end{figure*}

While manually classifying single-pulse candidates to use as a training set 
for our machine learning classifier (Section~\ref{sec:AI}), we identified a 
candidate Fast Radio Burst.  The burst (Figure~\ref{fig:0613+18}) was detected 
with $\rm DM = 400.3~{pc~cm}^{-3}$, width $W \approx 2~\rm ms$ and 
$\rm S/N = 8.4$ ($S_{\rm pk} = 39~\rm mJy$). The observed burst DM exceeds 
the Galactic maximum predicted along the line of sight ($\ell = 191\fdg9$, $b=+0\fdg36$) 
by both the NE2001 model 
\citep[$\rm DM_{\rm NE,max} = 188~{\rm pc~cm}^{-3}$, ][]{cl03} 
and the YMW16 model 
\citep[$\rm DM_{\rm YMW16,max} = 296~{\rm pc~cm}^{-3}$, ][]{ymw16}; we therefore classify 
it as a candidate FRB and refer to the burst as FRB~141113.  
To further investigate the reality of the event, we consider in detail both the significance of the burst detection 
and the robustness of the DM excess.

\subsection{Candidate Significance}
\label{ssec:frb_sig}
The false-alarm probability of a single-pulse detection at $\rm S/N = 8.4$ due
purely to Gaussian noise is vanishingly small.  In the presence of RFI,
however, statistical probabilities can be difficult to quantify reliably.  To
assess the significance of our detection of FRB~141113, we manually classified
{\it all} candidates in the database with $\rm S/N \geq 7$, 
$\rm DM = 300-3000\ pc\ cm^{-3}$ ($\rm DM = 2596\ pc\ cm^{-3}$ is the highest 
DM of all FRBs known to-date;~\citeauthor{bkb+18} \citeyear{bkb+18}) and 
$W \leq 10\ \rm ms$ (only 3 out of 30 FRBs have 
$W > 10~\rm ms$)\footnote{\label{frbcat}http://frbcat.org/}.  The manual 
classification was done by visually inspecting the single-pulse candidate (`spd') plot of each of 
the candidates that met our selection criteria and determining whether the 
candidate appeared astrophysical based on its frequency structure (broad-band 
and well described by a $\nu^{-2}$ law characteristic of cold plasma dispersion).  

\begin{figure*}[hbt]
    \centering
    \includegraphics[width=0.6\textwidth]{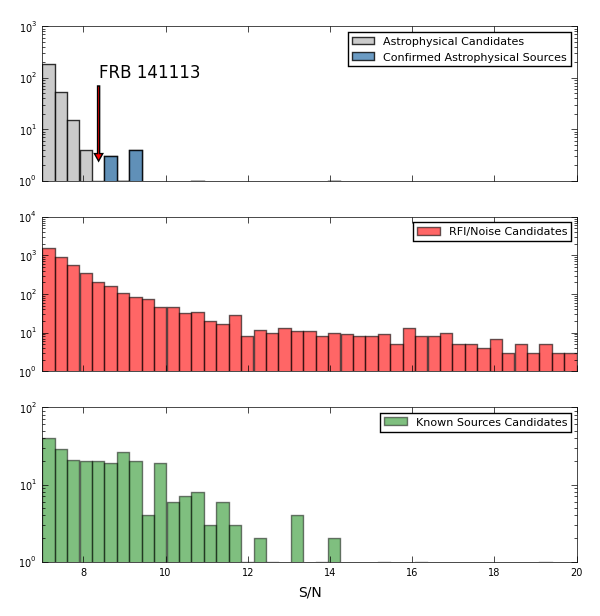}
    \caption{$\rm S/N$ distribution of candidates with $\rm S/N \geq 7$, 
             $\rm DM$ between $\rm 300\ pc\ cm^{-3}$ and 
             $\rm 3000\ pc\ cm^{-3}$ and pulse width $\rm \leq 10\ ms$ 
             as manually classified by members of PALFA collaboration. 
             \textit{Top:} Potential astrophysical candidates (totaling $\approx$270) 
             showing that FRB 141113 uniquely stands out from the population. 
             \textit{Middle:} Candidates classified as RFI or noise 
             (totaling $\approx$4,450). 
             \textit{Bottom:} Single pulses from known sources (totaling $\approx$270).  
             Note that there are multiple events for most known sources and that the vertical scale is not the same in each panel.}
    \label{fig:SNR_hist}
\end{figure*}

A distribution of the $\approx 5000$ manually classified candidates as a 
function of $\rm S/N$ is shown in Figure~\ref{fig:SNR_hist}. The top panel 
shows the distribution of all the $\approx 270$ candidates classified as likely 
astrophysical (some of which have already been confirmed as astrophysical via 
re-observations) by members of the collaboration. The middle panel shows the 
distribution of the $\approx 4500$ candidates classified as RFI or noise, and 
the bottom panel shows the distribution of $\approx 270$ pulses from known 
astrophysical sources.

Bright potential astrophysical candidates look qualitatively very different from
RFI (or noise) candidates and so are reliably classified.  However, weaker
potentially astrophysical signals are harder to distinguish from noise. Such
candidates are conservatively classified as noise. 
The distribution of candidates from known sources is relatively flat compared
to the other distributions because most of the known sources have multiple single pulse
candidates which span a wide range of $\rm S/N$. 

Importantly, {\it all} candidates with $\rm S/N > 8$ classified as potential
astrophysical sources have indeed been confirmed as pulsars or RRATs via
re-observations, except FRB~141113.  We henceforth assume the source to be astrophysical.


\subsection{Galactic DM Contribution}
\label{ssec:dm_excess}
%

We next consider whether candidate FRB 141113 is extragalactic, i.e. a genuine FRB, or whether
its excess DM could be caused by an intervening Galactic source not 
accounted for in the electron density models.

\subsubsection{Multiwavelength View of FRB Region}
\label{sssec:overview}
\begin{figure*}[h]
\centering
\includegraphics[width=0.45\textwidth]{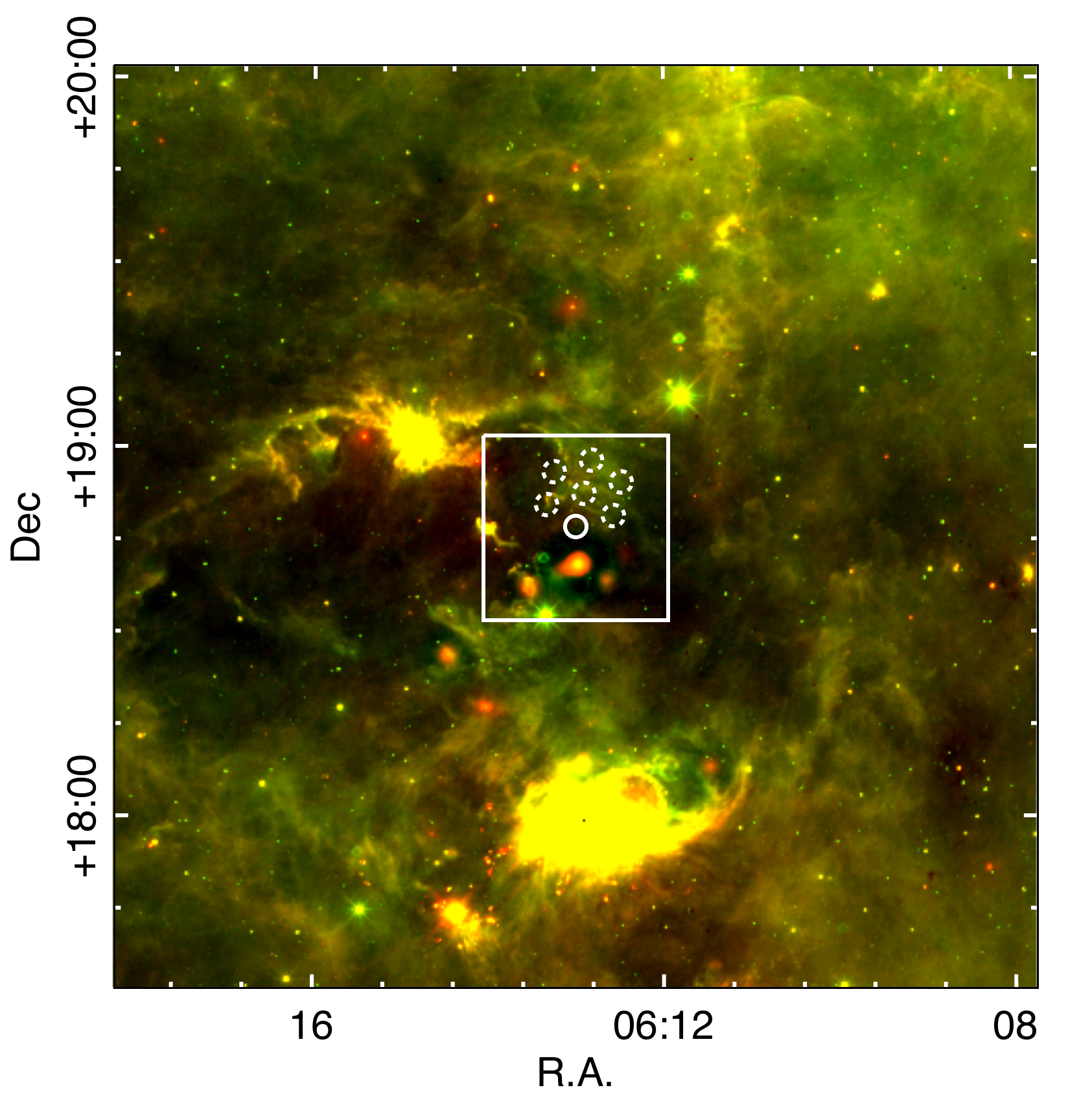}
\hspace{-2em}
\includegraphics[width=0.45\textwidth]{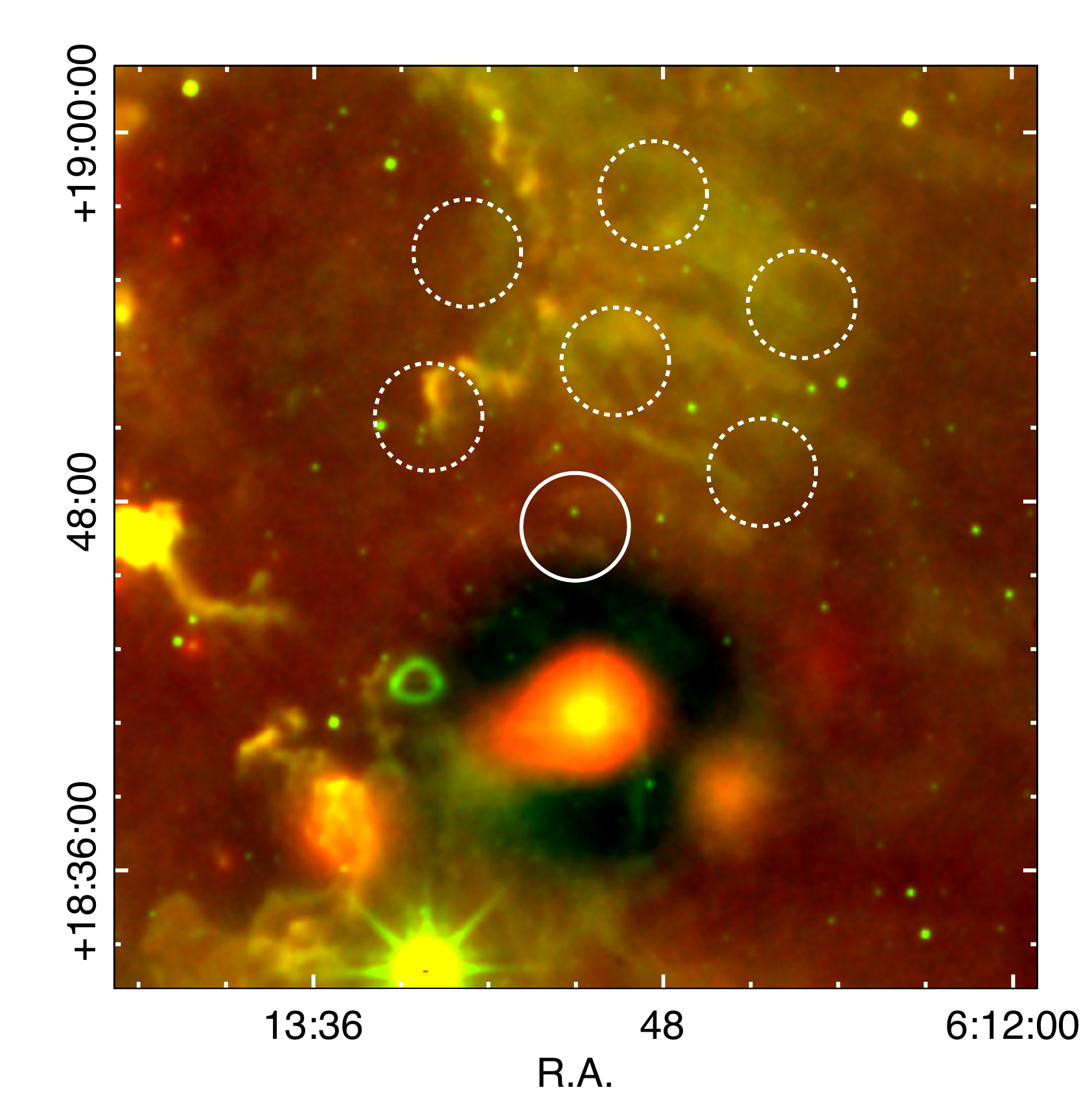}

\vspace{-3em}
\includegraphics[width=0.45\textwidth]{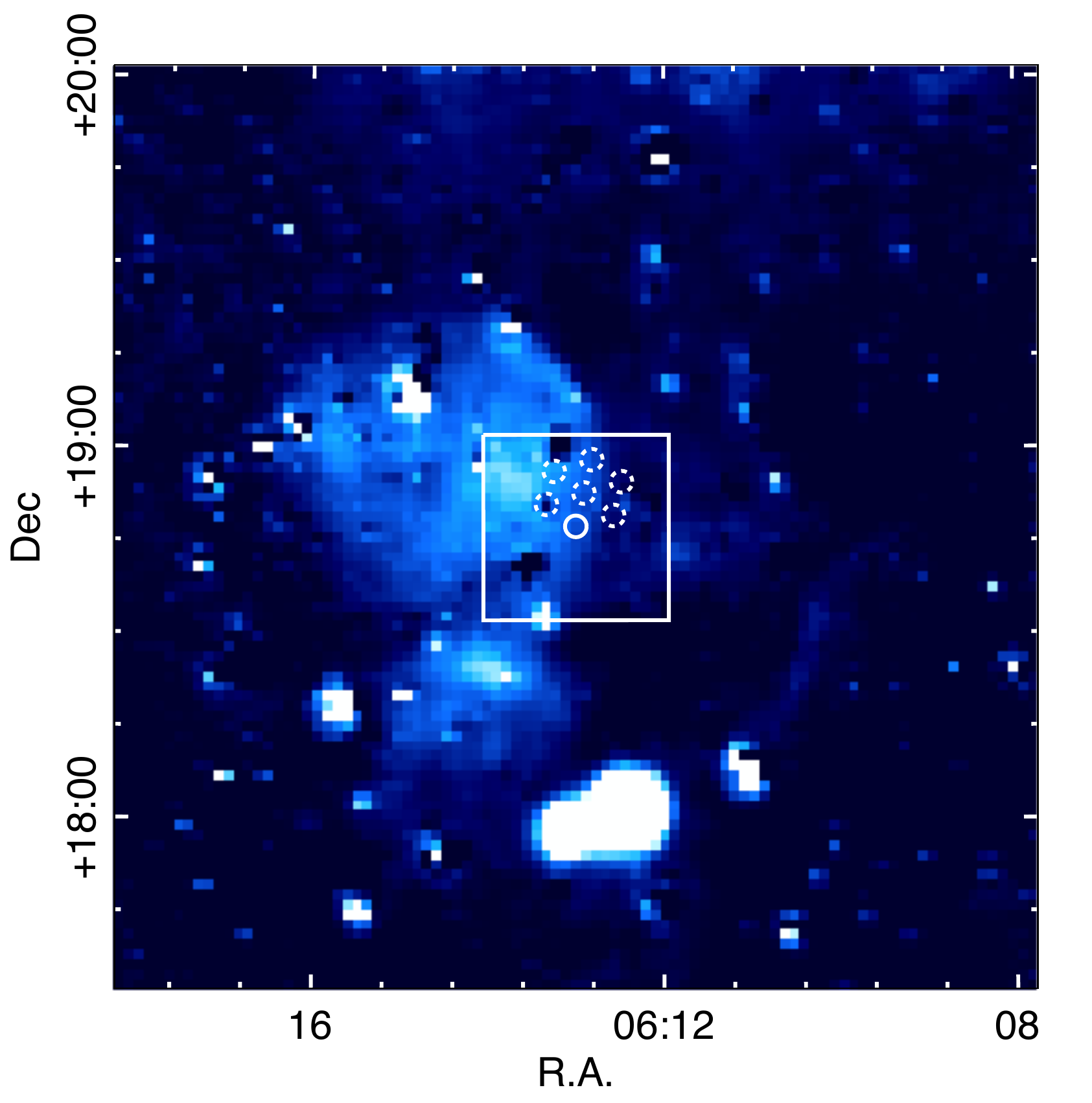} 
\hspace{-2em}
\includegraphics[width=0.45\textwidth]{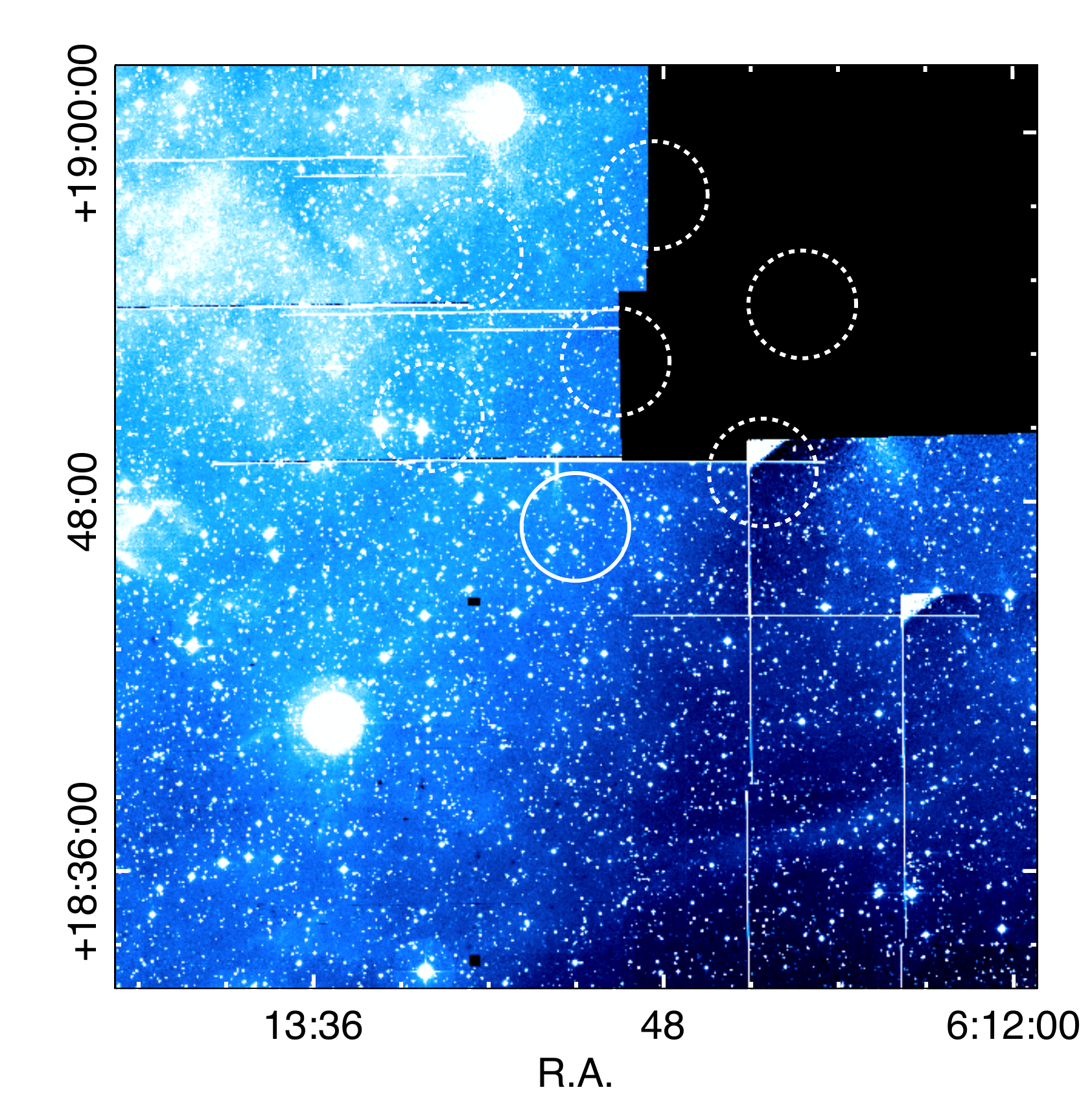} 

\vspace{-3em}
\includegraphics[width=0.45\textwidth]{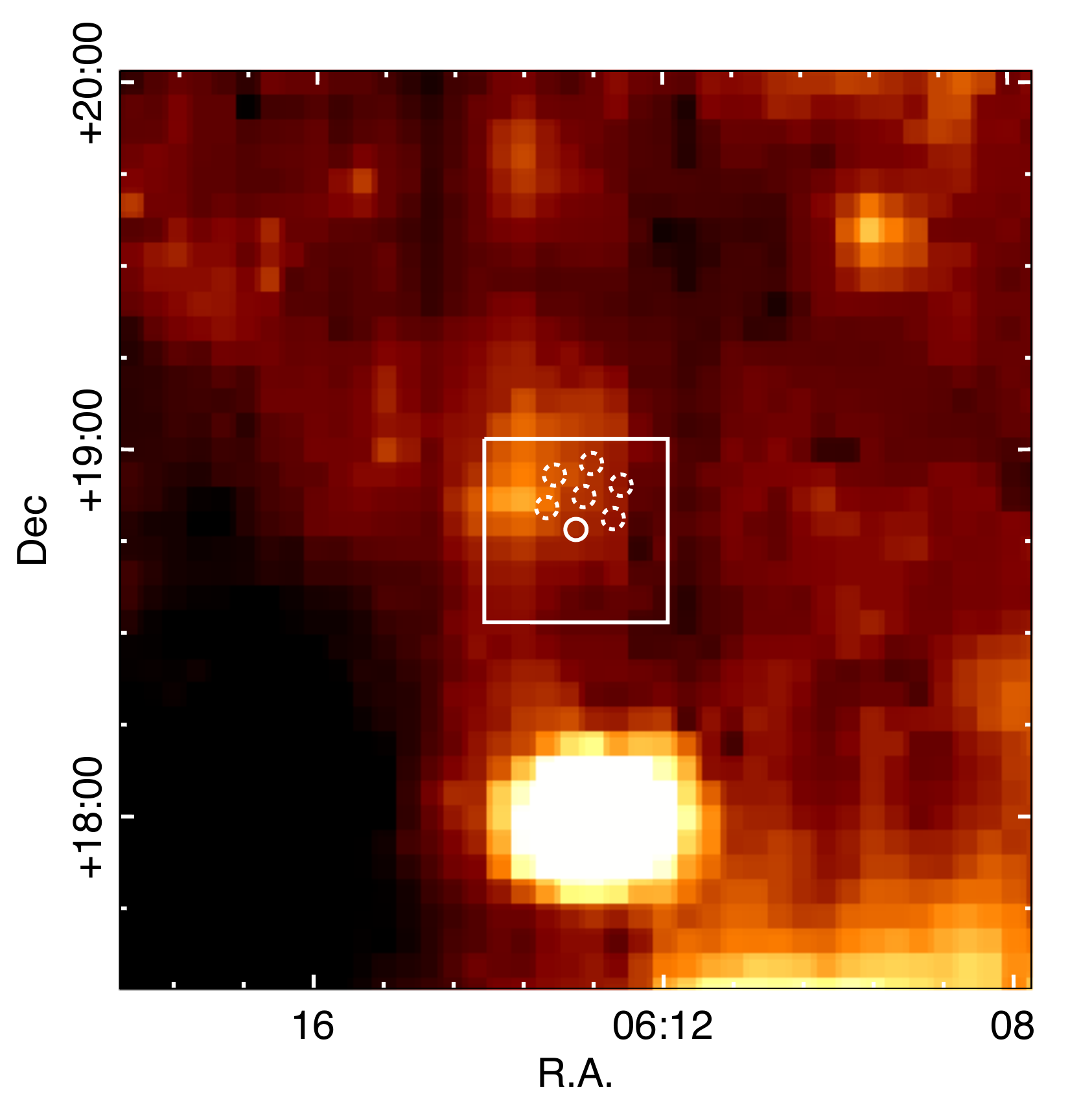}
\hspace{-2em}
\includegraphics[width=0.45\textwidth]{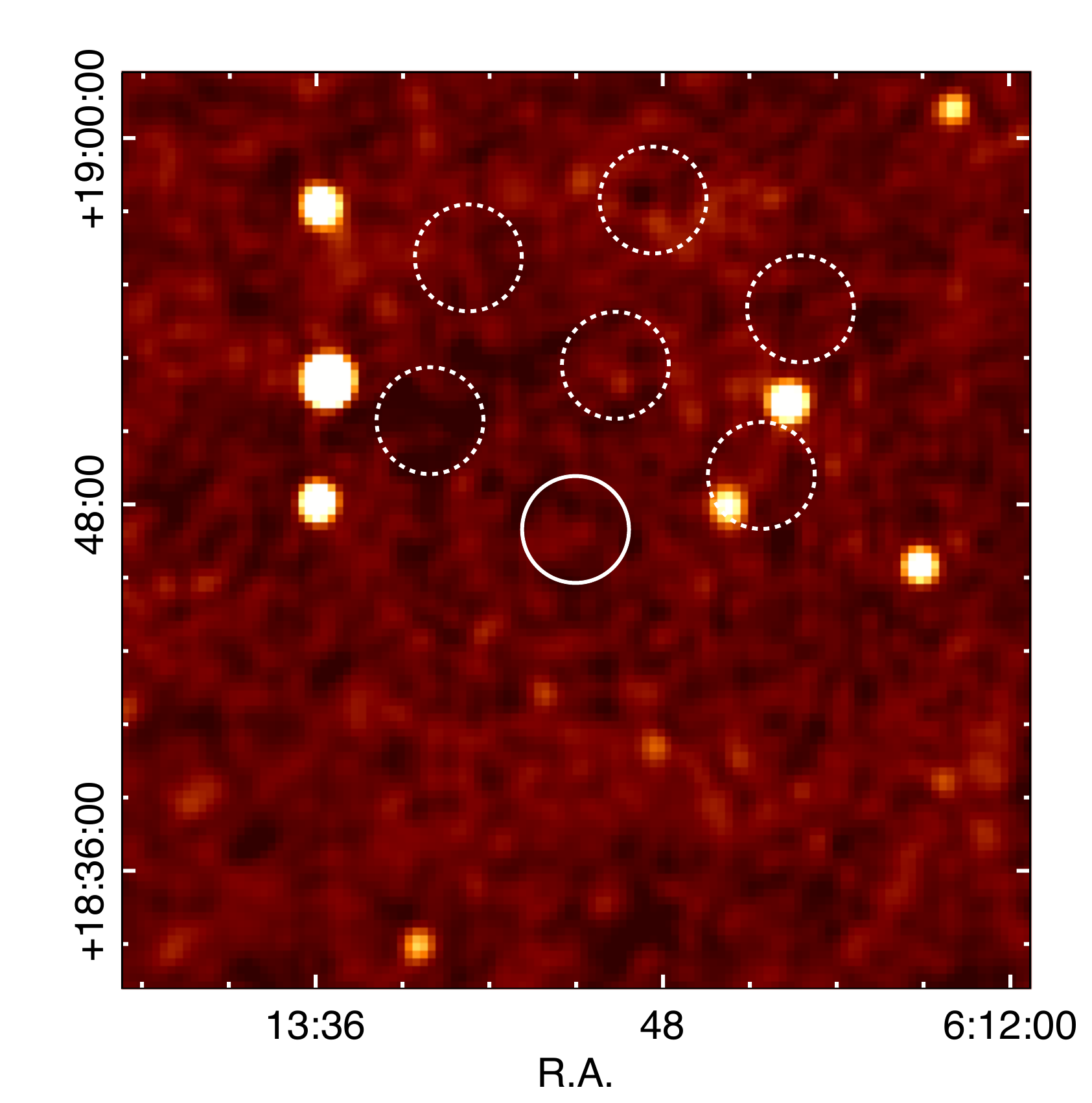}

\caption{FRB~141113 field in infrared, H$\alpha$, and 1.4~GHz radio bands.  
         The seven PALFA beams with $\rm HPBW = 3\farcm5$ are shown 
         (detection beam with solid line). The left column shows a $2\fdg5$ 
         square patch of sky centered on the detection beam position and the 
         right column shows the $30\arcmin$ region indicated by the white 
         square in the left column.  The top panels show WISE 
         $12\,\mu\rm m$ (green) and $22\,\mu\rm m$ (red). The center panels 
         show H$\alpha$ data from VTSS (left) and IPHAS (right).  The bottom 
         panels show 1.4~GHz radio maps from CHIPASS (left) and NVSS (right).  }
\label{fig:skyplots}
\end{figure*}

We search for $\rm H\textsc{ii}$ regions along the line of sight to FRB~141113 on 
angular scales from $\lesssim 1\arcsec$ to $\approx 1^{\circ}$ using 
both archival multi-wavelength data and a new VLA observation.  
Figure~\ref{fig:skyplots} shows the FRB field on two angular 
scales ($2\fdg5$ and $30\arcmin$) in the mid-infrared 
(useful to search for $\rm H\textsc{ii}$ regions), 
$\rm H \alpha$ (a tracer of ionized gas), and 
1.4~GHz radio (for free-free emission) bands.  The seven 
$\theta_{\rm FHWM} = 3\farcm5$ PALFA beams are shown in each panel 
with the detection beam indicated by a solid circle.

The mid-infrared panels of Figure~\ref{fig:skyplots} show 
$12~\mu \rm m$ (green) and $22~\mu \rm m$ (red) data from 
the WISE survey \citep{wem10}.  
In the $2\fdg5$ image, there are 
several structures with nebular morphology, notably a well known 
complex of $\rm H\textsc{ii}$ regions \citep[S254-258, ][]{cah08} 
about $45\arcmin$ south of the FRB detection beam and another 
about $20\arcmin$ to the east.  Most (if not all) of these regions 
lie within the Gemini~OB1 molecular cloud complex at a distance 
of $d \approx 2~\rm kpc$ from the Sun \citep{css95}.  In the 
$30\arcmin$ image, there is a bright imaging artifact 
$\approx 5\arcmin$ south, but no obvious $\rm H \textsc{ii}$ regions 
near the detection beam.

The $\rm H\alpha$ panels of Figure~\ref{fig:skyplots} show data from 
the Virginia Tech Spectral-line Survey \citep[VTSS, ][]{dst93} in the 
$2\fdg5$ field and IPHAS \citep{dgi05} in the $30\arcmin$ field. The 
VTSS image clearly shows the S254-258 $\rm H \textsc{ii}$ regions.  
It also shows a large ($\theta \approx 0\fdg8$ diameter) 
faint ($I_{\rm H\alpha} \approx 10-20~\rm R$\footnote{
$1~\rm R = 10^6/4\pi~{\rm photons}~{\rm cm}^{-2}~{\rm s}^{-1}~{\rm sr}^{-1}$}) 
structure that just barely overlaps the detection beam.  The 
IPHAS $30\arcmin$ image shows that while the brightest regions 
are to the north-east, there is still an elevated $\rm H\alpha$ flux 
coincident with the detection beam.

The 1.4~GHz radio panels of Figure~\ref{fig:skyplots} show data 
from the Parkes CHIPASS map \citep{csb14} in the $2\fdg5$ field and 
VLA NVSS data \citep{ccg98} in the $30\arcmin$ field.  The CHIPASS 
map shows an increase in the full-beam ($\theta_{\rm HPBW} = 14\arcmin$) 
brightness temperature of $\Delta T_{\rm b} \approx 100-200~\rm mK$ 
($S \approx 0.2-0.5~{\rm Jy~beam}^{-1}$ with $G = 0.44~{\rm K~Jy}^{-1}$) 
at roughly the same position as the $\rm H\alpha$ peak.  From the 
NVSS map, however, we see that a Parkes beam at this location would 
contain three point sources with total flux density of 
$S_{\rm sum} \approx 0.2~{\rm Jy}$ accounting for the rise in flux in the CHIPASS map.  There are no sources seen within the 
detection beam in the NVSS map.  

\begin{figure*}[hbt]
\begin{center}
\includegraphics[width=0.5\textwidth]{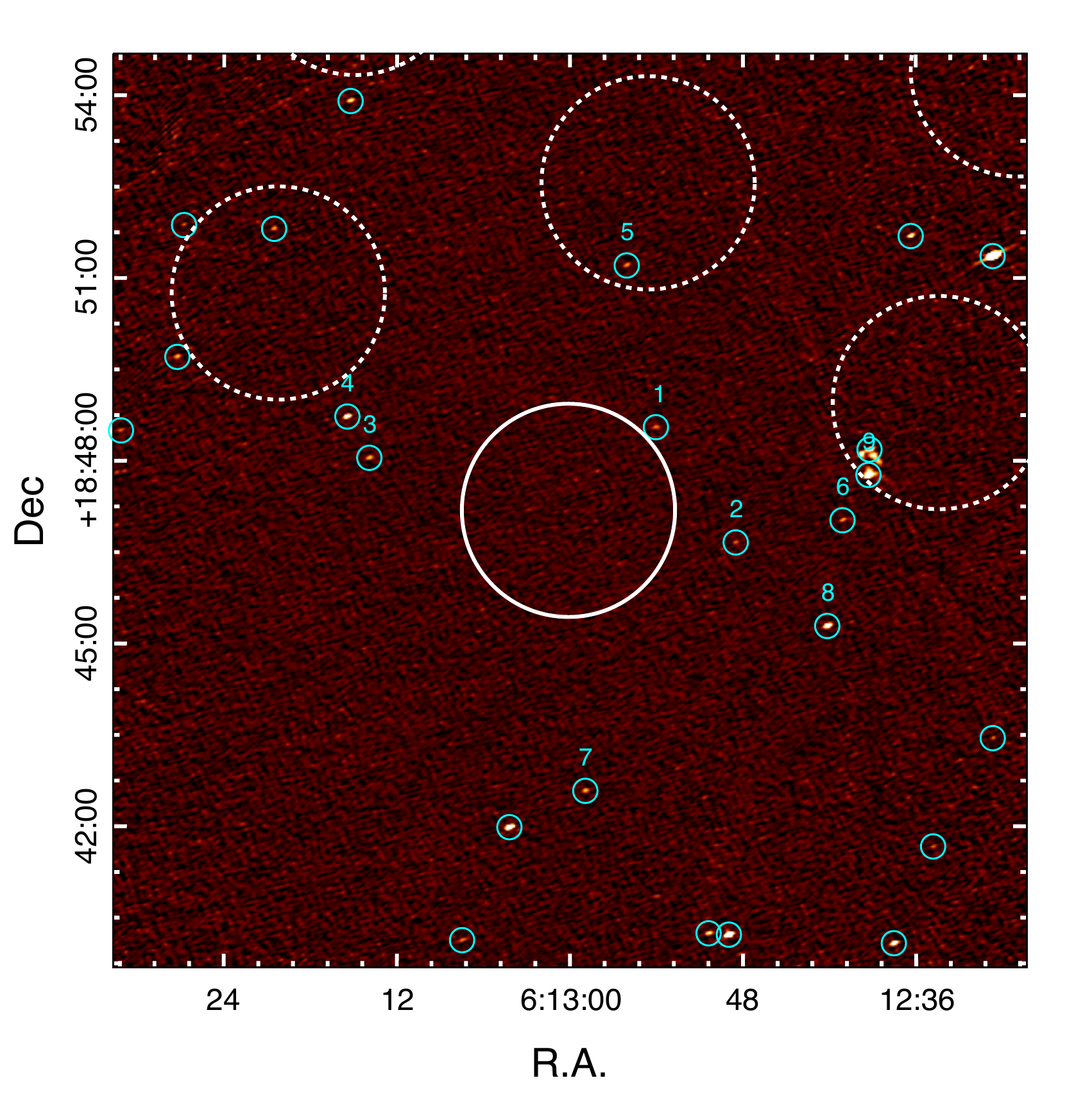}
\end{center}
   \caption{A $15\arcmin \times 15\arcmin$ VLA map of the FRB~141113 
            detection region at 1-2~GHz.  The solid white circle 
            shows the PALFA burst detection beam ($\theta_{\rm HPBW} = 3\farcm5$) 
            and the dashed circles show the other beams.  The cyan 
            circles show sources detected above a $5\sigma$ threshold of 
            $S_{\rm det} = 150~\rm \mu Jy$.  Sources within $5\arcmin$ of the 
            center of the detection beam are numbered in order of increasing 
            angular separation from the detection beam.  
            }
\label{fig:vla}
\end{figure*}

In addition to archival radio data, we also conducted observations with 
the Karl~G.~Jansky Very Large Array (VLA) to produce a sensitive radio map 
on arcsecond scales.  Observations were conducted on 2018 Jan~22 (MJD~58140) 
at 1-2~GHz with the array in B-configuration and resulted in about 12 minutes 
of time on source.  The absolute flux density calibrator 3C138 and the phase 
calibrator J0534+1927 were used. The data were calibrated and flagged using the 
VLA calibration pipeline. Additional RFI flagging and self-calibration were 
done after the pipeline calibration to produce a final primary-beam corrected 
image (Fig.~\ref{fig:vla}) with RMS noise of $\sigma \approx 30~{\rm \mu Jy~beam^{-1}}$ in the center 
of beam, which is consistent with expectations. 

\begin{figure}[!b]
\begin{center}
\includegraphics[width=3.0in]{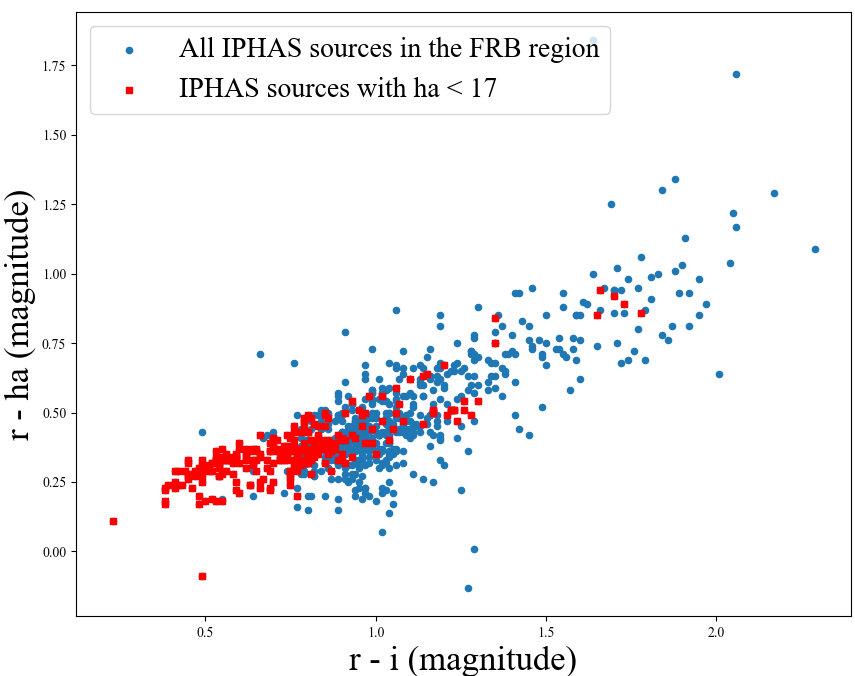}
\end{center}
   \caption{\small Color-color diagram for the 1135 IPHAS sources in the FRB region. 
            The sources shown as red squares $(ha < 17)$ are classified as either 
            stars or galaxies. Moreover, there is no source having $ha < 17$ lying above  
            cluster of points, consistent with none being an $\rm H\alpha$ emitter.}
\label{fig:iphas}
\end{figure} 

\subsubsection{DM Contribution from a Galactic Nebula?}
\label{sssec:nebula}
The excess DM FRB~141113,  
$\Delta {\rm DM} = {\rm DM} - {\rm DM}_{\rm mod}$, 
is $\Delta {\rm DM}_{\rm NE} = 212~{\rm pc~cm}^{-3}$ for NE2001 and 
$\Delta {\rm DM}_{\rm YMW16} = 104~{\rm pc~cm}^{-3}$ for YMW16.  We explore 
whether there could exist an unmodelled Galactic ionized region contributing this excess
along the line of sight to FRB~141113, and set limits in various wavebands on any relevant emission.  

A hypothetical homogeneous spherical nebula along the FRB sight-line with 
${\Delta \rm DM} = 212~{\rm pc~cm}^{-3} = n_{\rm e} L_{\rm pc}$ 
($L_{\rm pc}$ is the nebula size in parsecs) would have an emission 
measure of at least 
${\rm EM} = {\rm \Delta DM}^{2} / L_{\rm pc} = 45000~{\rm pc~cm}^{-6}~ L_{\rm pc}^{-1}$.  
For an electron temperature of 8000~K, the nebula has a free-free optical 
depth of 
\begin{equation}
  \tau_{\rm ff} = \frac{0.01}{L_{\rm pc}} \, 
                \left(\frac{T_{\rm e}}{\rm 8000~K}\right)^{-1.35} \, 
                \left(\frac{\nu}{\rm 1.4~GHz}\right)^{-2.1}.
\label{eq:tauff}
\end{equation}
Since an optically thick nebula is rendered implausible by the absence
of a detection of any ultra-compact H\textsc{ii} region in the WISE H\textsc{ii} 
region survey \citep{anderson2014wise}, we require that the nebula be optically 
thin.  This requirement sets a lower limit of $L_{\rm min} = 0.01~\rm pc$ on the 
size of the nebula.  Introducing a filling factor in the expressions for 
$\Delta \rm DM$  and EM only makes the limits below more constraining 
\citep{Kulkarni2014}. 

Using the IPHAS point source catalog \citep{Drew2005}, we search for 
$\rm H\alpha$ emission from a compact nebula.  
Following the method described in \cite{Kulkarni2015} and \cite{Scholz2016}, 
we estimate that the $\rm H\alpha$ flux for a compact nebula at 20~kpc 
in standard IPHAS magnitude units ($ha$) would be $ha < 17$.  This assumes 
$L_{\rm pc} = 0.01~{\rm pc}$, imposed by the optically thin condition 
(Eq.~\ref{eq:tauff}). From the IPHAS point source catalog, there are 
1135 catalogued sources in 5$'$-radius region around
the nominal FRB position,
out of which 159 objects have $ha < 17$. None is classified as an 
$\rm H\alpha$ emitter \citep{Barentsen2014}. Another method of classifying 
$\rm H \alpha$ emitters is with a color-color diagram \citep{Kulkarni2015}. 
For the FRB region, this is shown in Figure~\ref{fig:iphas}. Any sources lying above the cluster of points would be $\rm H\alpha$ emitter 
candidates. However, we see none  having $ha < 17$.  
Therefore, IPHAS strongly constrains the presence of an unresolved Galactic 
nebula in the FRB region. Assuming a larger nebula or closer distance would 
strengthen this conclusion.

\begin{figure*}[h]
\centering
\includegraphics[width=0.75\textwidth]{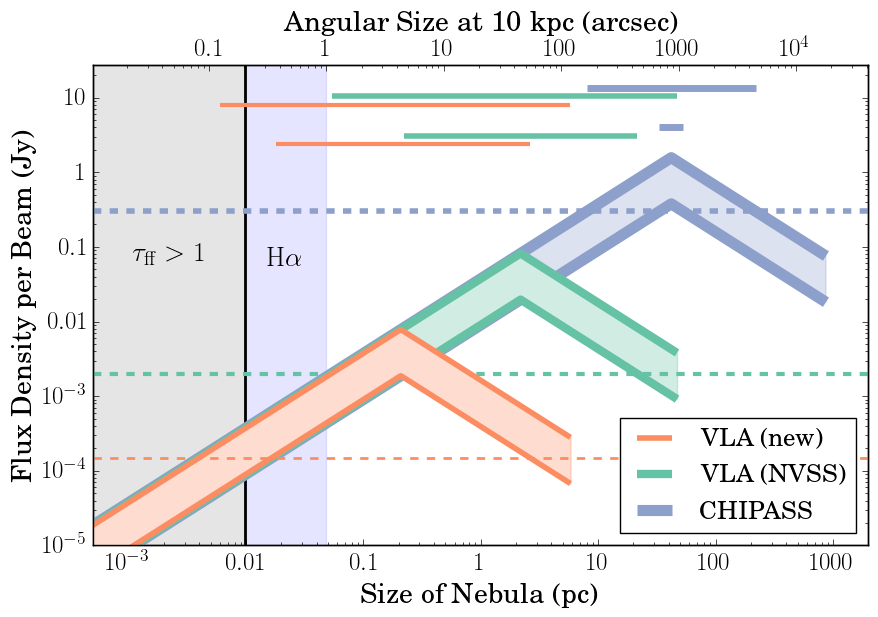}
\caption{Predicted flux density per beam from free-free emission 
         of a nebula at 10~kpc contributing the excess DM toward  
         FRB~141113 seen with recent VLA B-configuration observations 
         (orange), the VLA NVSS (green), and Parkes CHIPASS (purple). 
         In each case, the upper and lower lines correspond to the 
         excess DM relative to the NE2001 and YMW16 electron density 
         models, respectively. 
         The dashed horizontal lines indicate the observed upper limit 
         for each survey.  
         The solid horizontal bars indicate nebula sizes that are 
         excluded by observation for the NE2001 (upper) and YMW16 (lower) 
         electron density models.
         The shaded regions are excluded due to the 
         requirement that the plasma be optically thin (grey) and 
         non-detections from IPHAS $\rm H\alpha$ point sources (blue).  
         }
\label{fig:FRBradio}
\end{figure*}

Next, we consider the free-free emission from a nebula that contributes 
the excess DM seen toward FRB~141113.  Following \citet{Scholz2016}, 
we calculate the 1.4~GHz flux density as a function of nebula size 
(Figure~\ref{fig:FRBradio}).  
In order to cover a wide range of angular sizes, we use data from our newly 
observed VLA B-configuration observations ($\theta_{\rm B} = 4\arcsec$), 
archival VLA NVSS data ($\theta_{\rm NVSS} = 45\arcsec$), and single dish 
Parkes data from CHIPASS ($\theta_{\rm P} =  14\farcm4$).  

At the largest angular scales, we set a limit of 
$S_{\rm max} = 0.3~\rm Jy$ in the Parkes beam 
($\rm HPBW = 14\farcm4$) as discussed in Section~\ref{sssec:overview}.
At smaller angular scales, we can use VLA observations.  In the 
NVSS map ($\theta_{\rm HPBW} = 45\arcsec$, 
$\sigma = 0.4~{\rm mJy~beam}^{-1}$), there are no sources detected 
above $5\sigma$ in the PALFA burst detection beam 
($\theta_{\rm HPBW} = 3\farcm5$). The nearest detected source is 
$5\arcmin$ from the center of the burst detection beam and falls 
within another PALFA beam (Figure~\ref{fig:skyplots}).  We set a 
$5\sigma$ upper limit of $S_{\rm max} = 2~{\rm mJy~beam}^{-1}$ 
from the NVSS map.

In the VLA B-configuration map ($\theta_{\rm HPBW} = 4\arcsec$, 
$\sigma = 30~{\rm \mu Jy~beam}^{-1}$), there are no sources detected 
above $5\sigma$ in the PALFA burst detection beam.  Searching out to 
$\Delta \theta = 5\arcmin$ (a distance that would roughly include the 
sidelobes of one PALFA beam), we find nine sources (Fig.~\ref{fig:vla}).
Five are located closer to a PALFA beam in which there was no detection, so we exclude these from consideration.  None of 
the four remaining (Sources 1, 2, 7, 8) is reported as having $\rm H\alpha$ 
emission in the IPHAS point source catalog.  Hence, these are unlikely to be HII regions. We set 
a $5\sigma$ upper limit of $S_{\rm max} = 150~{\rm \mu Jy~beam}^{-1}$ 
from the VLA B-configuration map.


\subsubsection{Galactic or Extragalactic?}
\label{sssec:origin}
Figure~\ref{fig:FRBradio} shows the predicted free-free emission from 
a nebula at 10~kpc accounting for the DM excess $\Delta {\rm DM}$ 
seen toward FRB~141113 for both the NE2001 and YMW16 electron density 
models.  Using limits from the free-free radio emission, $\rm H\alpha$ 
emission, and the optically thin plasma requirement, we exclude nebulae 
with angular sizes of up to several degrees out to 10~kpc assuming the 
NE2001 excess.  The smaller DM excess predicted by the YMW16 model allows 
for the possibility that an extended ($\theta \gtrsim 15\arcmin$) and 
distant ($d \geq 8~\rm kpc$) nebula contributes the observed DM excess.  
An extended ($\theta \approx 0\fdg8$) source is seen in the $\rm H\alpha$ 
image (Figure~\ref{fig:skyplots}), but it is almost certainly associated 
with the Gemini OB1 molecular cloud complex at $d\approx 2~\rm kpc$ and 
would be ruled out by the free-free limits.  Overall, we conclude that 
FRB~141113 is very likely extragalactic.

\subsection{Host Galaxy and IGM Contribution}
\label{ssec:host}
Assuming now that FRB~141113 is genuine and extragalactic, we calculate 
the host and IGM contributions to the observed DM as 
$\rm DM_{IGM} + DM_{host} = DM - DM_{\rm NE,max} - DM_{halo} \approx 182~{pc~cm}^{-3}$
with $\rm DM_{halo} \approx 30~{pc~cm}^{-3}$ and 
$\rm DM_{\rm NE,max} =188~{pc~cm}^{-3}$. 
Using the $\rm DM_{IGM}$-redshift scaling relation 
${\rm DM} \approx 1200 z~{\rm pc~cm}^{-3}$ for $z \leq 2$ 
\citep{ioka2003cosmic,inoue2004probing}, we estimate a redshift of  
$z < 0.15$, which corresponds to a distance of approximately 0.6~Gpc. This, and the low flux 
of 39~mJy (Table~\ref{tab:discoveries}) suggest the FRB~141113 could be one of the 
closest FRBs with one of the lowest luminosities yet detected.  However, detection in 
a sidelobe, which would imply a much larger source flux, cannot presently be excluded.  
If the source repeats, an interferometric position determination will be possible, 
and the true source flux could be established.  Multi-wavelength follow-up would be
warranted given its relatively nearby location compared with other FRBs.
Monitoring observations are ongoing at the Arecibo Observatory.


\section{Implications for the FRB population}
\label{sec:implications}
We have found that FRB 141113 is likely to be a genuine extragalactic cosmic event.
An additional check on its authenticity is to verify whether its detection in PALFA is consistent with reported event rates and constraints on the flux density distribution of the FRB population. 

\subsection{FRB Detection Rate}\label{sec:rate}
The sensitivity of the PALFA survey allows detection of bursts in the FWHM region of the beam (hereafter, main beam) and in the near side-lobes, thus requiring characterization of both to determine the FRB rate. For the main beam, the field-of-view (FOV) is $\Omega = 0.022$ sq. deg, and the mean system flux S$_\mathrm{sys}$ = 5 Jy, and for the full FOV of 0.105 sq. deg., S$_\mathrm{sys}$ = 27 Jy. The full FOV includes the main beam and regions of the near side-lobes with gain greater than the Parkes 1.4-GHz average gain of 0.4 K/Jy \citep{sch+14}. Based on the above-mentioned system fluxes, S/N detection threshold of (S/N)$_\mathrm{b}$ = 8,
 $n_\textrm{p} = 2$ and $\Delta f = 322 \; \textrm{MHz}$, we estimate the minimum detectable flux densities for the main and full beams to be 44 mJy and 239 mJy, respectively. The calculation is performed using Equation 1 for an intrinsic pulse width of 3 ms assuming no scatter-broadening, and accounts for the degradation in sensitivity by a factor of 1.5, as discussed in \S\ref{sec:sensitivity}. Additionally, we adopt (S/N)$_\mathrm{b}$ = 8 instead of 7, which was employed in the sensitivity analysis in \S\ref{sec:sensitivity}, because of the ambiguity in determining whether a candidate with (S/N)$_\mathrm{b} < 8$ is RFI or astrophysical (see Fig.~\ref{fig:SNR_hist}). 

We adopt T$_\mathrm{obs} = 24.1$ days as an estimate of the total observation time for PALFA pointings processed by the modified analysis pipeline. The estimate is obtained after subtracting time corresponding to the mean masking fraction due to RFI of 10\%, assuming that all masking was done in the time domain. Additionally, pointings with masking fraction greater than 20\% were not processed by the pipeline and hence were not included in the estimate. Although scattering in the inner Galaxy can hinder FRB detection, over 97\% of the included pointings have predicted maximum scattering timescales of $<$2 ms along their line of sight, thus ensuring minimal effect on the results of the following analyses.

Based on the detection of one likely event (i.e. FRB 141113) in observations of 0.022 sq. deg. of sky for a duration of 24 days, we estimate the FRB detection rate for the main beam of the PALFA survey to be 7.8$^{+35.6}_{-7.6} \times \ 10^4$ FRBs sky$^{-1}$ day$^{-1}$ above a threshold of 44 mJy, with the 95\% confidence interval evaluated assuming Poisson statistics. Accounting for the possibility of the burst being detected in the near side-lobes as for the repeating FRB 121102 (\citealt{sch+14}; \citealt{chatterjee2017}), we estimate 1.6$^{+7.5}_{-1.6} \times \ 10^4$ FRBs sky$^{-1}$ day$^{-1}$ above a threshold of 239 mJy. The above estimate assumes uniform sensitivity to bursts with diverse spectral behavior, such as those detected from the repeating FRB 121102 \citep{Scholz2016}.

We have not updated the rate estimate
reported by \citet{Scholz2016} which was based on the detection of FRB 121102 in T$_\mathrm{obs}$ = 36.9 days. This is
because the estimate is derived from the results of an analysis pipeline with a sensitivity different from that of the pipeline described here. Although there is some overlap between data processed by the two, data pertaining to FRB 121102 have not yet
been processed by the modified pipeline. We note that our
reported rate is greater than the rate derived by \citet{Scholz2016}. However, the 95\% confidence intervals for both have substantial overlap, implying that the detection of candidate FRB
141113 is consistent with the \citet{Scholz2016} estimate.  

\subsection{Log $N$--Log $S$ Distribution}
The observed cumulative flux density distribution of the FRB population is modelled as a power law with an index $\alpha$ (hereafter, the log $N$--log $S$ slope) such that the number of FRBs with a flux density greater than $S$ is $N(>S) \propto S^{-\alpha}$. For a local, uniformly distributed, non-evolving source population, $\alpha$ = 1.5 with any deviation from this value supporting the existence of a cosmological and/or evolving source population.

Here we derive constraints on $\alpha$ by performing simulations of cumulative flux density distributions of the FRB population. These simulations utilize results from the analysis pipeline detailed in \citet{lbh+15} which searched T$_{obs}$ = 36.9 days with a threshold S/N of 9.2 (hereafter, search A) and the analysis pipeline discussed in this work searching T$_\mathrm{obs}$ = 24 days with a threshold S/N of 8 (hereafter, search B). Observations from these two searches are key in constraining the log $N$--log $S$ slope.  We include the near side-lobe detection of FRB 121102 in search A and assume, at least initially, that FRB 141113 was detected in the main beam for search B.  We also account for non-detections in the main beam and near side-lobes, for searches A and B, respectively, under our initial assumption. The sensitivity threshold and sky coverage assumed for the main beam are discussed in Section \ref{sec:rate} while those for the near side-lobe are calculated based on the corresponding values for the full FOV after subtracting the contribution of the main beam. Additionally, we account for the non-detection of any event in the far-out side-lobes for both these searches. Although the survey is sensitive to such ultra-bright off-axis bursts occurring over the visible hemisphere with a sensitivity described by Equation 19 of \citet{dcm+09}, their occurrence can likely be ruled out due to absence of multi-beam detections. 

The simulations were performed by varying the log $N$--log $S$ slope in the range, $0 < \alpha \leq 2$, in steps of 0.1. All trial values were assumed to be equally probable with thousands of runs performed for each. For each of these runs, a flux density distribution was generated which was consistent with the low-latitude FRB rate of 285$^{+1416}_{-237}$ bursts sky$^{-1}$ day$^{-1}$ above 1 Jy, estimated by \citet{vanderwiel2016}. Based on these flux density distributions, we computed a detection rate $R$, in bursts sky$^{-1}$ day$^{-1}$, above the sensitivity thresholds corresponding to the main beam, as well as the near and far-out side-lobes for both searches A and B. The number of detections for a given search and ALFA beam region for each simulation run is sampled from a Poisson distribution with a mean of $R T_\mathrm{obs}\Omega$, where $\Omega$ is defined in \S\ref{sec:rate}. A run is counted as a success if the number of simulated detections for all regions of the ALFA beam for both searches is equal to that for the observations. An additional criterion for a successful run is the flux density of the detected bursts in the simulations lying in the range of possible flux densities for the observed bursts (FRB 121102 and candidate FRB 141113).

For determining flux densities of the observed bursts, we injected pulses with DM and widths equal to those of FRB 121102 and candidate FRB 141113 in PALFA pointings and obtained the range for which these pulses are detected with the same S/N as observed in the pipeline. The system flux used in this analysis varied for the two sources. The mean system flux for the main beam of 5 Jy was used for FRB 141113 as it is not possible to localize the burst position in the ALFA beam. However, we can obtain a better estimate for the gain and hence the system flux for the position of FRB 121102 as it has been localized to milliarcsecond precision owing to its repeat bursts \citep{chatterjee2017}. We model the ALFA beam pattern \citep{sch+14} and find the gain at the position of FRB 121102 to be 0.6--0.7 K/Jy (accounting for ALFA pointing errors) which we use to calculate the system flux and the observed flux density. 

\begin{figure}[h]
\centering
\includegraphics[scale=0.61]{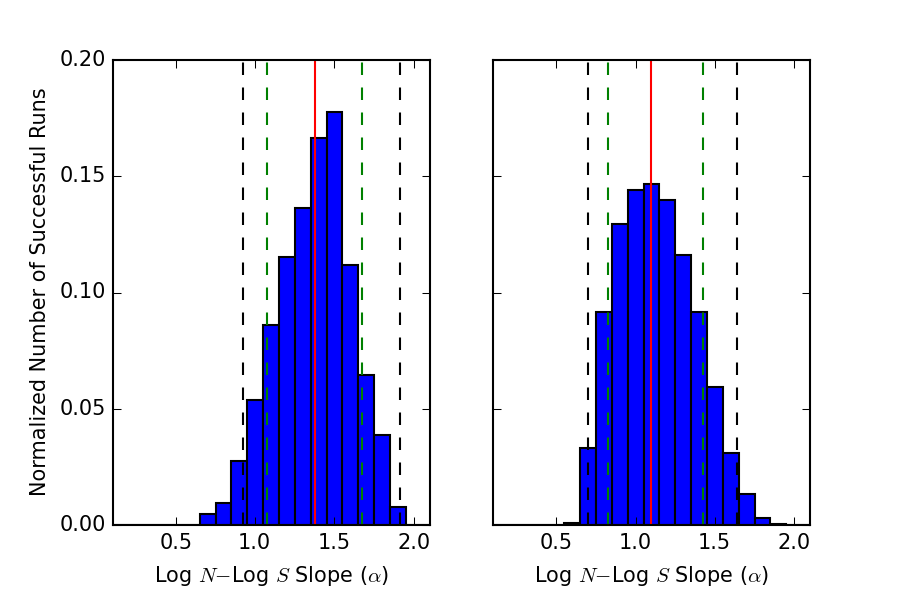}
\caption{Normalized number of MC runs for which the number and flux density of detections matched with results of FRB searches with the PALFA survey, plotted for all trial values of $\alpha$. While constraints on $\alpha$ in both panels are based on the detection of FRB 121102, the left panel further assumes that FRB 141113 is astrophysical, while the right panel is for it being a false positive. The median value for $\alpha$ is denoted by the red, solid line with 1$\sigma$ and 2$\sigma$ confidence intervals denoted by the green and black dashed lines, respectively.} 
\label{fig:constraint}
\end{figure}

Based on the relative number of successful runs for each trial value of $\alpha$, normalized by the total number of runs and plotted in the left panel of Figure \ref{fig:constraint}, we find that the detection of candidate FRB 141113 and additional PALFA observations imply a median $\alpha$ of 1.4 with the 95\% confidence interval ranging from 0.9 $< \alpha <$ 1.9. We reject $\alpha < 0.9$ at the 95\% confidence level since the implied abundance of bright bursts is inconsistent with the lack of off-axis multi-beam detections with PALFA. Steeper log $N$--log $S$ slopes ($\alpha > 1.9$) are rejected since detection of a single faint burst is unlikely considering the implied abundance of faint bursts in this case.

The above constraint is on the {\it observed} log $N$--log $S$ slope, which due to propagation effects, can be different from the slope intrinsic to the population. While diffractive interstellar scintillation with its small decorrelation bandwidth at low Galactic latitudes is unlikely to be important \citep{macquart2015}, effects such as plasma lensing in FRB host galaxies can enhance flux densities of faint bursts \citep{cordes2017}. 

Our above reported constraints have substantial overlap with those reported for the observed log $N$--log $S$ slope by \citet{oppermann2016} ($0.8 < \alpha < 1.7$), \citet{caleb2016} ($0.6 < \alpha < 1.2$) and \citet{lawrence2017} ($0.57 < \alpha < 1.25$). However, these constraints are inconsistent with those reported by \citet{vedantham2016} (0.5 $< \alpha <$ 0.9) based on multiple-beam detections with Parkes surveys and other detections with telescopes of varied diameters. By running our simulations for the case of the candidate FRB 141113 being a false positive event, we find a significant shift in our constraints to a median $\alpha$ of 1.1 with 95\% bounds ranging from 0.7 to 1.6 (see right panel of Fig. \ref{fig:constraint}), which has overlap with the \citet{vedantham2016} constraints. Confirming whether the event is an FRB by observations of repeat bursts could thus have strong implications for studies of the cumulative flux density distribution of the FRB population.

Additionally, our constraints are in tension with those estimated by \citet{bkb+18} ($1.6 < \alpha < 3.4$) and \citet{macquart2018} ($1.9 < \alpha < 3.9$) using a maximum likelihood analysis technique for FRBs detected with the Parkes telescope above the observationally complete fluence threshold of 2 Jy ms. Such steep log $N$--log $S$ slopes predicting an abundance of faint bursts are already unlikely based on the event rate implied by the discovery of FRB 121102 with the PALFA survey \citep{Scholz2016}. However, constraints based on results from the Arecibo and Parkes telescopes can be reconciled if the log $N$--log $S$ slope flattens at low flux densities in which case a single power law cannot describe the flux density distribution of the observed FRB population, as suggested by \citet{macquart2018}.

Our reported constraints depend strongly on our assumptions. Varying the reference FRB rate to be the all-sky estimate of 587$^{+337}_{-315}$ FRBs sky$^{-1}$ day$^{-1}$ above a peak flux density of 1 Jy reported by \citet{lawrence2017} yields $\alpha = 1.2^{+0.5}_{-0.4}$ (95\% bounds). Additionally, assuming FRB 141113 to have been in the near side-lobe instead of the main beam modifies the constraint to be $\alpha = 1.25^{+0.5}_{-0.4}$.\footnote{We do not consider the possibility of the candidate FRB 141113 being an off-axis detection as it is difficult to know the fraction of the field of view for which particular ray paths into the optics of the ALFA receiver could result in a single-beam detection. Therefore, our reported constraints might not be valid if the flux density of the candidate FRB was greater than $\sim$10$^5$ Jy.  However, this is unlikely considering that no bursts brighter than 9.2 KJy were detected in a search at 1.4-GHz conducted with the Bleien Radio Observatory for an observing time of 590 days \citep{sainthilaire2014}.} Although there are factors we did not account for while calculating the range of fiducial gain values for FRB 121102 (for e.g., rotation of the receiver at the time of observation), we find no significant change in our constraints even if the full range of gains possible for the inner edge of the side-lobe of ALFA is used (0.4--1.0 K/Jy; \citealt{sch+14}).

\section{Conclusion}
\label{sec:conclusion}
We have described a new, more systematic single-pulse pipeline to improve the search for pulsars, RRATs, and FRBs in the PALFA survey. The pipeline adds post-processing features to efficiently identify astrophysical single pulses. 

We also performed a robust sensitivity analysis of the PALFA survey to single pulses using injection of synthetic signals into survey data. 
We find that for pulse widths $\rm < 5\ ms$ our survey is at most a factor of $\rm \sim 2$ less sensitive to single pulses than the theoretical predictions. For pulse widths $\rm > 10\ ms$, as the $\rm DM$ decreases, the degradation in sensitivity gets worse by up to a factor of $\rm \sim 4.5$.
In order to better understand the actual sensitivities to single pulses in various radio transient surveys, we recommend similar characterization of their deployed detection pipelines.

Using our pipeline, we have discovered one pulsar and two RRATs that were not detected using periodicity searching techniques, six pulsars that were detected by both single pulse and periodicity pipelines, three candidate RRATs, and one candidate FRB. 
This latter source, FRB 141113, 
has a DM more than twice the likely Galactic maximum along the line of sight, and multi-wavelength observations show it is very likely to be extragalactic.  If so, it is consistent with being one of the lowest luminosity FRBs yet discovered. Simulations accounting for the sensitivity of PALFA and the discovery of FRB 121102 in addition to this new source indicate that the slope of the log $\rm N$--log $\rm S$ relation for the FRB population (i.e., $N(>S) \propto S^{-\alpha}$) is $\alpha = 1.4 \pm 0.5$ (95\% confidence).  The steepness of that distribution is at odds with previous suggestions of a much flatter slope \citep{vedantham2016}.  However, relaxing some reasonable assumptions in our calculation results in somewhat lower mean slopes, with uncertainty ranges that still bracket flatter population distributions.

\section*{Acknowledgements}
DA is supported by the NSF OIA-1458952.
MB is supported by a Mitacs Globalink Graduate Fellowship. AB, FC, SC, JMC, SMR, IHS, MAM, DRL, WWZ, RF, BN, and KS are members of the NANOGrav Physics Frontiers Center, which is supported by the National Science Foundation award number 1430284. PC is supported by an FRQNT Doctoral Research Award and a Mitacs Globalink Graduate Fellowship. EP is a Vanier Scholar. VMK holds the Lorne Trottier Chair in Astrophysics \& Cosmology and a Canada Research Chair and receives support from an NSERC Discovery Grant and Herzberg Award, from an R. Howard Webster Foundation Fellowship from the Canadian Institute for Advanced Research (CIFAR), and from the FRQNT Centre de Recherche en Astrophysique du Quebec. DRL is also supported by the NSF AST-1516958 and NSF OIA-1458952. PS is supported by a DRAO Covington Fellowship from the National Research Council Canada. MAM acknowledges support from the and NSF Award Number 1458952. The National Radio Astronomy Observatory is a facility of the National Science Foundation operated under cooperative agreement by Associated Universities, Inc.  SMR is a CIFAR Senior Fellow. RSW acknowledges financial support by the European Research Council (ERC) for the ERC Synergy Grant BlackHoleCam under contract no. 610058. WWZ is supported by the CAS Pioneer Hundred Talents Program, National Key R\&D Program of China No. 2017YFA0402600 and National Nature Science Foundation of China No. 11743002. KCV acknowledges the following ARCC students who have contributed to observations: Brent Cole, Keith Bohler and Yhamil Garcia. JSD is supported by the NASA Fermi program. PCCF gratefully acknowledges financial support by the European Research Council for the Starting grant BEACON under contract No. 279702, and continued support from the Max Planck Society. JWTH acknowledges funding from an NWO Vidi fellowship and from the European Research Council under the European Union's Seventh Framework Programme (FP/2007-2013) / ERC Starting Grant agreement nr. 337062 ("DRAGNET").
Pulsar research at UBC is supported by an NSERC Discovery Grant and by CIFAR. The research leading to these results has received funding from the European Research Council under the European Union's Seventh Framework Programme (FP/2007-2013) / ERC Grant Agreement n. 617199. The Arecibo Observatory is operated by SRI International under a cooperative agreement with the National Science Foundation (AST-1100968), and in alliance with Ana G.M\'{e}ndez-Universidad Metropolitana, and the Universities Space Research Association. The CyberSKA project was funded by a CANARIE NEP-2 grant. Computations were made on the supercomputer Guillimin at McGill University, managed by Calcul Qu\'{e}bec and Compute Canada. The operation of this supercomputer is funded by the Canada Foundation for Innovation (CFI), NanoQu\'{e}bec, RMGA and the Fonds de recherche du Qu\'{e}bec$-$Nature et technologies (FRQ-NT).

\bibliography{references,devansh,pragya,FRB,rsw}
\bibliographystyle{aasjournal}

\end{document}